\documentclass[twocolumn,aps,superscriptaddress,showpacs]{revtex4-1}

\usepackage{mathptmx}
\usepackage{dcolumn}
\usepackage{amsmath,amssymb,amsfonts}
\usepackage{dsfont}
\usepackage{bm}
\usepackage{color}
\usepackage{latexsym}
\usepackage{epstopdf}
\usepackage[english]{babel}
\usepackage{enumerate}
\definecolor{mygrey}{gray}{0.35}
\definecolor{myblue}{rgb}{0.2,0.2,0.8}
\definecolor{myzard}{cmyk}{0,0,0.05,0}
\definecolor{mywhite}{rgb}{1,1,1}
\definecolor{mywhite}{rgb}{1,1,1}
\definecolor{myred}{rgb}{1,0.,0.3}

\usepackage[pdftex]{graphicx}

\usepackage{natbib}
\usepackage{epstopdf}
\usepackage{appendix}

\def\be{\begin{equation}}
\def\ee{\end{equation}}
\def\ba{\begin{align}}
\def\enda{\end{align}}
\def\bi{\begin{itemize}}
\def\ei{\end{itemize}}

 \def\Z{{\mathbb{Z}}}

 \def\ee{\mathord{\rm e}}

\def\min{\mathord{\rm min}}

\renewcommand{\ee}{{\rm e}}
\renewcommand{\aa}{{\rm a}}

\def\beq{\begin{equation}}
\def\eeq{\end{equation}}

 \newcommand{\ket}[1]{|#1\rangle}
 \newcommand{\bra}[1]{\langle #1|}

\def\LL{{\rm L}}

\def\PP{\mathbb{P}}
\def\QQ{\mathbb{Q}}

\newcommand{\ud}[1]{{#1^{\dagger}}}
\newcommand{\braket}[2]{\langle #1|#2\rangle}

\newcommand{\mean}[1]{\langle #1\rangle}

\newcommand{\kk}{\mathbf{k}}
\newcommand{\rr}{\mathbf{r}}

\begin{document}

\title{Quantum Spin Dynamics with Pairwise-Tunable, Long-Range Interactions}

 \author{C.-L. Hung}
   \email[Corresponding author:]{clhung@purdue.edu}
 \affiliation{Department of Physics and Purdue Quantum Center, Purdue University, West Lafayette, IN 47907, USA}
  
 \author{A. Gonz\'{a}lez-Tudela}
  \email[Corresponding author:]{alejandro.gonzalez-tudela@mpq.mpg.de}
 \affiliation{Max-Planck-Institut f\"{u}r Quantenoptik Hans-Kopfermann-Str. 1.
85748 Garching, Germany }
      
\author{J. I. Cirac}
 \affiliation{Max-Planck-Institut f\"{u}r Quantenoptik Hans-Kopfermann-Str. 1.
85748 Garching, Germany }

 \author{H. J. Kimble}
  \affiliation{Max-Planck-Institut f\"{u}r Quantenoptik Hans-Kopfermann-Str. 1.
85748 Garching, Germany }
 \affiliation{Norman Bridge Laboratory of Physics 12-33}
  \affiliation{Institute for Quantum Information and Matter, California Institute of Technology, Pasadena, CA 91125, USA}
 
\date{\today}

\begin{abstract}
We present a platform for the simulation of quantum magnetism with full control of interactions between pairs of spins at arbitrary distances in one- and two-dimensional lattices. In our scheme, two internal atomic states represent a pseudo-spin for atoms trapped within a photonic crystal waveguide (PCW). With the atomic transition frequency aligned inside a band gap of the PCW, virtual photons mediate coherent spin-spin interactions between lattice sites. To obtain full control of interaction coefficients at arbitrary atom-atom separations, ground-state energy shifts are introduced as a function of distance across the PCW. In conjunction with auxiliary pump fields, spin-exchange versus atom-atom separation can be engineered with arbitrary magnitude and phase, and arranged to introduce non-trivial Berry phases in the spin lattice, thus opening new avenues for realizing novel topological spin models. We illustrate the broad applicability of our scheme by explicit construction for several well known spin models.
\end{abstract}

\maketitle

\section*{Introduction}
Quantum simulation has become an important theme for research in contemporary physics~\cite{cirac12a}. A quantum simulator consists of quantum particles (e.g., neutral atoms) that interact by way of a variety of processes, such as atomic collisions. Such processes typically lead to short-range, nearest-neighbor interactions~\cite{jaksch05a, bloch12a, trotzky2008, simon2011, greif2013}. Alternative approaches for quantum simulation employ dipolar quantum gases, polar molecules, and Rydberg atoms~\cite{griesmaier05a, micheli06a, lu11a, jaksch00a,ni08a, saffman10a}, leading to interactions that typically scale as $1/r^3$, where $r$ is the inter-particle separation. For trapped ion quantum simulators \cite{porras04a, blatt12a, islam13}, tunability in a power law scaling of $r^{-\eta}$ with $0< \eta<3$ can in principle be achieved. Beyond simple power law scaling, it is also possible to engineer arbitrary long-range interactions mediated by the collective phonon modes, which can be achieved by independent Raman-addressing on individual ions \cite{korenblit12}. 

Using photons to mediate controllable long-range interactions between isolated quantum systems presents yet another novel approach for assembling quantum simulators \cite{kimble08a}. Recent successful approaches include coupling ultracold atoms to a driven photonic mode in a conventional mirror cavity, thereby creating quantum many-body models (using atomic external degrees of freedom) with cavity-field mediated infinite-ragne interactions \cite{baumann10}. Finite-range and spatially disordered interactions can be realized by employing multi-mode cavities \cite{gopalakrishnan09}. Recent demonstrations on coupling cold atoms to guided mode photons in photonic crystal waveguides \cite{goban13a, goban15a} and cavities \cite{thompson13a, tiecke14a} present promising new avenues (using atomic internal degrees of freedom) due to unprecedented strong single atom-photon coupling rate and scalability. Related efforts also exists for coupling solid-state quantum emitters, such as quantum dots \cite{majumdar12a, javadi15} 
and diamond nitrogen-vacancy centers \cite{barclay09, hausmann13}, to photonic crystals. Scaling to a many-body quantum simulator based on solid-state systems, however, still remains elusive. Successful implementations can be found in the microwave domain, where superconducting qubits behave as artificial atoms strongly coupled to microwave photons propagating in a network formed by superconducting resonators and transmission lines \cite{Houck2012,eichler2015,Mckay2015}.

\begin{figure}
\centering
\includegraphics[width=0.8\columnwidth,keepaspectratio]{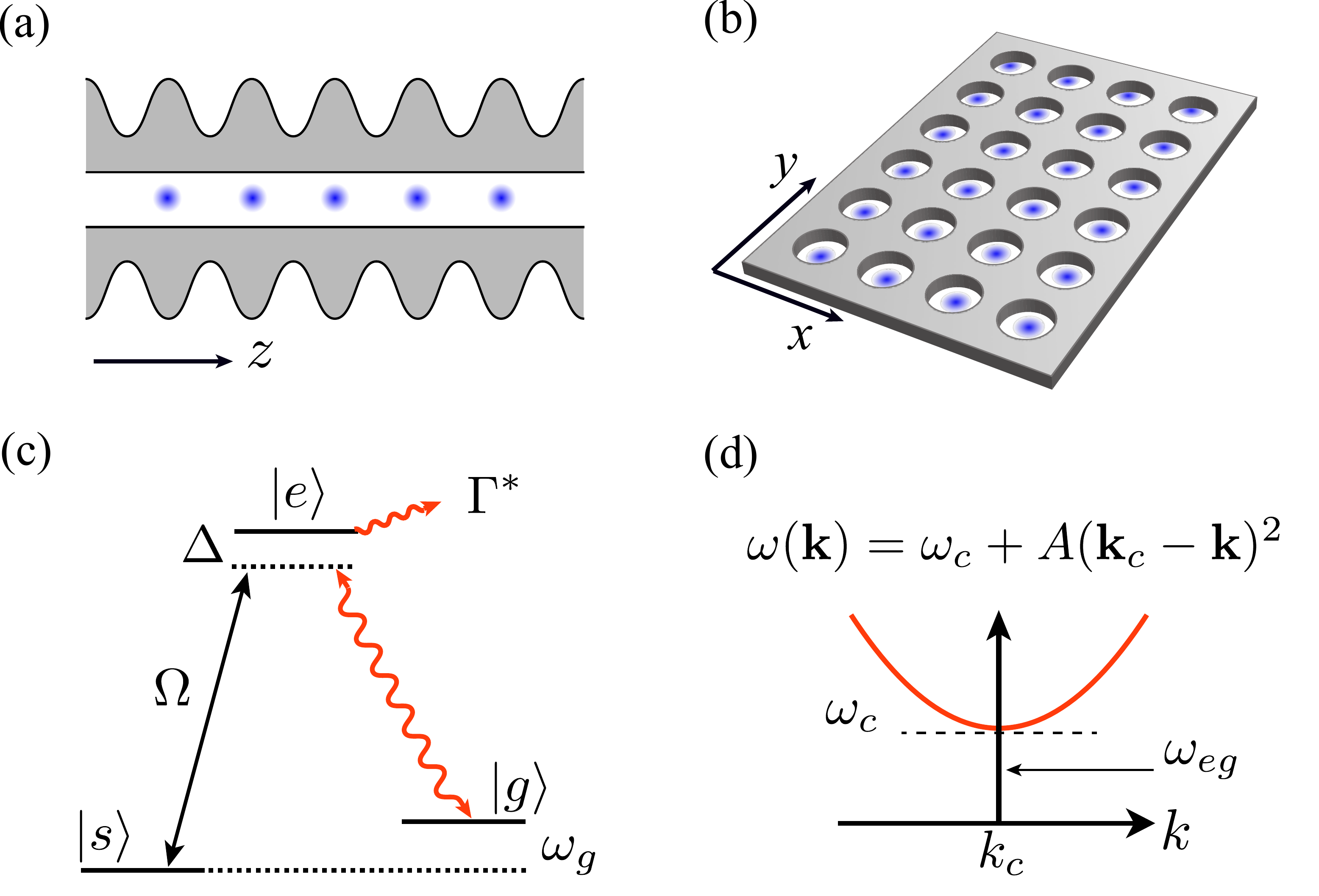}
\caption{Photon-mediated atom-atom interactions in (a) 1D and (b) 2D PCWs. (c) Atomic level scheme: atomic dipole $\ket{s} \leftrightarrow \ket{e}$ is coupled to an external pump, $\ket{g}\leftrightarrow \ket{e}$ coupled to a GM photon, and $\Gamma^*$ the excited state decay rate to free space and leaky modes. (d) Simplified band structure $\omega(\kk)$ near the band edge $\mathbf{k}=\mathbf{k}_c$ and $\omega(\mathbf{k}_c) = \omega_c$. Atomic transition frequency $\omega_{eg}=\omega_e-\omega_g$ lies within the band gap.}
\label{fig1}
\end{figure}

Here we propose and analyze a physical platform for simulating long-range quantum magnetism in which full control is achieved for the spin-exchange coefficient between a pair of spins at arbitrary distances in one- (1D) and two-dimensional (2D) lattices. The enabling platform, as described in Refs.~\cite{douglas15a,gonzaleztudela15c}, is trapped atoms within photonic crystal waveguides (PCWs), with atom-atom interactions mediated by photons of the guided modes (GMs) in the PCWs. As illustrated in Fig.~\ref{fig1}(a-b), single atoms are localized within unit cells of the PCWs in 1D and 2D periodic dielectric structures. At each site, two internal atomic states are treated as pseudo-spin states, with spin-1/2 considered here for definiteness [e.g., states $\ket{g}$ and $\ket{s}$ in Fig.~\ref{fig1}(c)].

Our scheme utilizes strong, and coherent atom-photon interactions inside a photonic band gap \cite{kurizki90a, john90a, john91a, douglas15a,gonzaleztudela15c}, and long-range transport property of GM photons for the exploration of a large class of quantum magnetism. This is contrary to conventional hybrid schemes based on, e.g., arrays of high finesse cavities \cite{hartmann06, greentree06a, cho08a, hartmann08a} in which the pseudo-spin acquires only the nearest (or at most the next-nearest) neighbor interactions due to strong exponential suppression of photonic wave packet beyond single cavities. 

In its original form \cite{kurizki90a, john90a, john91a, douglas15a,gonzaleztudela15c}, the localization of pseudo-spin is effectively controlled by single-atom defect cavities \cite{douglas15a}. The cavity mode function can be adjusted to extend over long distances within the PCWs, thereby permitting long-range spin exchange interactions. The interaction can also be tuned dynamically, via external addressing beams, to induce complex long-range spin transport, which we describe in the following \cite{douglas15a,gonzaleztudela15c}. 

To engineer tunable, long-range spin Hamiltonians, we use an atomic $\Lambda$ scheme and two-photon Raman transitions, where an atom flips its spin state by scattering one photon from an external pump field into the GMs of a PCW. The guided-mode photon then propagates within the waveguide, inducing spin flip in an atom located at a distant site via the reverse two-photon Raman process. When we align the atomic resonant frequency inside the photonic band gap, as depicted in Fig.~\ref{fig1}(d), only virtual photons can mediate this remote spin exchange and the GM dynamics are fully coherent, effectively creating a spin Hamiltonian with long-range interactions. As discussed in Refs.~\cite{douglas15a,gonzaleztudela15c}, the overall strength and length scale of the spin-exchange coefficients can be tuned by an external pump field, albeit within the constraints set by a functional form that depends on the dimensionality and the photonic band structure. 

To fully control spin-exchange coefficients at arbitrary separations, here we adopt a Raman-addressing scheme similarly discussed for cold atoms in optical lattices \cite{jaksch03, aidelsburger13, miyake13}. We introduce atomic ground-state energy shifts as a function of distance across the PCW. Due to conservation of energy, these shifts suppress reverse two-photon Raman processes in the original scheme \cite{douglas15a,gonzaleztudela15c}, forbidding spin exchange within the entire PCW. However, we can selectively activate certain spin-exchange interactions $J(r_{m,n})$ between atom pairs $(m,n)$ separated by $r_{m,n}$, by applying an auxiliary sideband whose frequency matches that of the original pump plus the ground state energy shift between the atom pairs. This allows us to build a prescribed spin Hamiltonian with interaction terms ``one by one''. Note that each sideband in a Raman-addressing beam can be easily introduced, e.g., by an electro-optical modulator. By introducing multiple sidebands and by 
controlling their frequencies, 
amplitudes and relative phases, we can engineer spin 
Hamiltonians with arbitrary, complex interaction coefficients $J(r_{m,n})$. Depending on the dimensionality and the type of spin Hamiltonians, our scheme requires only one or a few Raman beams to generate the desired interactions.
Furthermore, by properly choosing the propagation phases of the Raman beams, we can imprint geometric phases in the spin system, thus opening new possibilities for realizing novel topological spin models.

We substantiate the broad applicability of our methods by explicit elaboration of the set of pump fields required to realize well-known spin Hamiltonians. For one-dimensional spin chains, we consider the implementation of the Haldane-Shastry model \cite{haldane88a,shastry88a}.
For two-dimensional spin lattices, we elaborate the configurations for realizing topological flat-bands \cite{neupert11a, sun11a} in Haldane's spin model \cite{haldane88a}, as well as a ``checkerboard'' chiral-flux lattice \cite{neupert11a, sun11a}. We also consider a 2D XXZ spin Hamiltonian with $J(r_{m,n}) \propto1/r_{m,n}^{\eta}$ and $\eta=1,2,3$ \cite{maik12a}. In addition, we report numerical results on the $\eta$-dependence of its phase diagram.

\section*{Controlling spin-spin interaction through multi-frequency driving}

In the following, we discuss how to achieve full control of interactions by multi-frequency pump fields. We assume (i) $N$ atoms trapped in either a 1D or 2D PCW as depicted in Fig.\ref{fig1}(a-b). For simplicity, we assume one atom per unit cell of the PCW, although this assumption can be relaxed afterwards; (ii) The structure is engineered \cite{thompson13a,goban13a,tiecke14a,goban15a} such that the GM polarization is coupled to the atomic dipole, $\ket{g}\leftrightarrow\ket{e}$, as shown in Fig.~\ref{fig1}(c) and, under rotating wave approximation, is described by the following Hamiltonian (using $\hbar=1$) 
\begin{equation}
 H_{\mathrm{lm}}=\sum_{\kk,n}g_\kk (\rr_n)a_\kk \sigma_{eg}^n+\mathrm{h.c.}\,,
\end{equation}
where $g_\kk (\rr_n)=g_\kk e^{i \kk\cdot \rr_n}$ is the single-photon coupling constant at site location $\rr_n$, with $n$ being the site index, $a_\kk$ the GM field operator, and atomic operators $\sigma_{\alpha\beta}^n\equiv\ket{\alpha}_n\bra{\beta}$.  Moreover, as in Refs.~\cite{douglas15a,gonzaleztudela15c}, we assume (iii) there is another hyperfine level $\ket{s}$, addressed by a Raman field with coupling strength $\Omega$ as follows  
\begin{equation}
 H_{\mathrm{d}}(t)=\sum_n\big(\frac{\Omega(t)}{2}\sigma_{se}^n e^{i\omega_{L} t}+\mathrm{h.c.}\big)\,,\label{eq:driving}
\end{equation}
where $\omega_L$ is the main driving frequency. The Raman field $\Omega(t)$ contains $m_P$ frequency components that are introduced to achieve full control of the final effective spin Hamiltonian. Full dependence of $\Omega(t)$ can be written as
\begin{equation}
 \label{eq:muli}
 \Omega(t)\equiv  \sum_{\alpha=0}^{m_p-1}\Omega_\alpha e^{i\tilde{\omega}_\alpha t}\,,
\end{equation}
where $\tilde{\omega}_\alpha$ are the detunings of the sidebands from the main frequency $\omega_\mathrm{L}$ such that $\tilde{\omega}_0=0$, and $\Omega_\alpha$'s are the complex amplitudes.

We can adiabatically eliminate the excited states $\ket{e}$ and the photonic GMs under the condition that (iv) $\mathrm{max}\{|\Omega|, |\tilde{\omega}_\alpha-\tilde{\omega}_\beta|\}\ll |\Delta|=|\omega_e-\omega_\LL|$. This condition guarantees that, firstly, the excited state is only virtually populated, and that, secondly, the time-dependence induced by the sideband driving is approximately constant over the timescale $\Delta^{-1}$. As discussed in Refs. \cite{douglas15a,gonzaleztudela15c}, if $\omega_L-\omega_{g}$ lies in the photonic band gap, photon-mediated interactions by GMs are purely coherent. \footnote{To simplify the discussion, in this manuscript we neglect decoherence effects caused by atomic emission into free space and leaky modes as well as photon loss due to imperfections in the PCW. These effects were both carefully discussed in Refs.~\cite{douglas15a,gonzaleztudela15c}.} Under the Born-Markov approximation, we then arrive at an effective XY Hamiltonian (SI Appendix \ref{secSM:derivation})
,
\begin{align}
 \label{eq:XYham}
 H_{XY}(t)=\sum_{m,n\neq m}^N\sum_{\alpha,\beta=0}^{m_P-1}X_{\alpha}X_{\beta}^*\tilde{J} e^{i(\omega_{g,m}-\omega_{g,n}+\tilde{\omega}_{\alpha}-\tilde{\omega}_{\beta})t}\sigma_{gs}^m\sigma_{sg}^n\,,
\end{align}
where we have defined $X_\alpha=\Omega_\alpha/(2\Delta)$; $\omega_{g,n} = \omega_g(\rr_{n})$ is the site dependent ground state energy shift, and $\tilde{J}(\rr_{m,n})$ is the atom-GM photon coupling strength \cite{douglas15a,gonzaleztudela15c} that typically depends on atomic separation $\rr_{m,n}=\rr_m -\rr _n$. 

We focus on ``sideband engineering'', and treat $|\tilde{J}(\rr_{m,n})|$ as approximately constant over atomic separations considered \footnote{One may also replace a PCW with a single-mode nanophotonic cavity, operating in the \emph{strong dispersive} regime \cite{blais04a,schuster07a}, to achieve constant GM coupling $\tilde{J}$ independent of $|\rr_{m,n}|$. Realistic nanophotonic cavity implementations will be considered elsewhere.}. This is valid as long as $|\rr_{m,n}|$ is much smaller than the decay length scale $\xi$ of the coupling strength $\tilde{J}(\rr_{m,n})$. Specifically, we require $|\rr_{m,n}| \ll \xi = \sqrt{\left|A/\Delta_c\right|}$, where $A$ is the band curvature, $\Delta_c =\mathrm{max}\{ \omega_c - (\omega_\LL-\omega_{g,n})\}$ is the maximal detuning of the band edge to the frequency of coupled virtual photons that mediate interactions [see Fig.~\ref{fig1}(c)], and we have assumed that the variation of ground state energies $\omega_ {g,n}$ are small compared to $\Delta_c$. Exact 
functional form of $ \tilde{J}(\rr_{m,n})$ can be found in Refs. \cite{douglas15a,gonzaleztudela15c} and in SI Appendix~\ref{secSM:derivation}. 

\begin{figure}
\centering
\includegraphics[width=0.8\columnwidth,keepaspectratio]{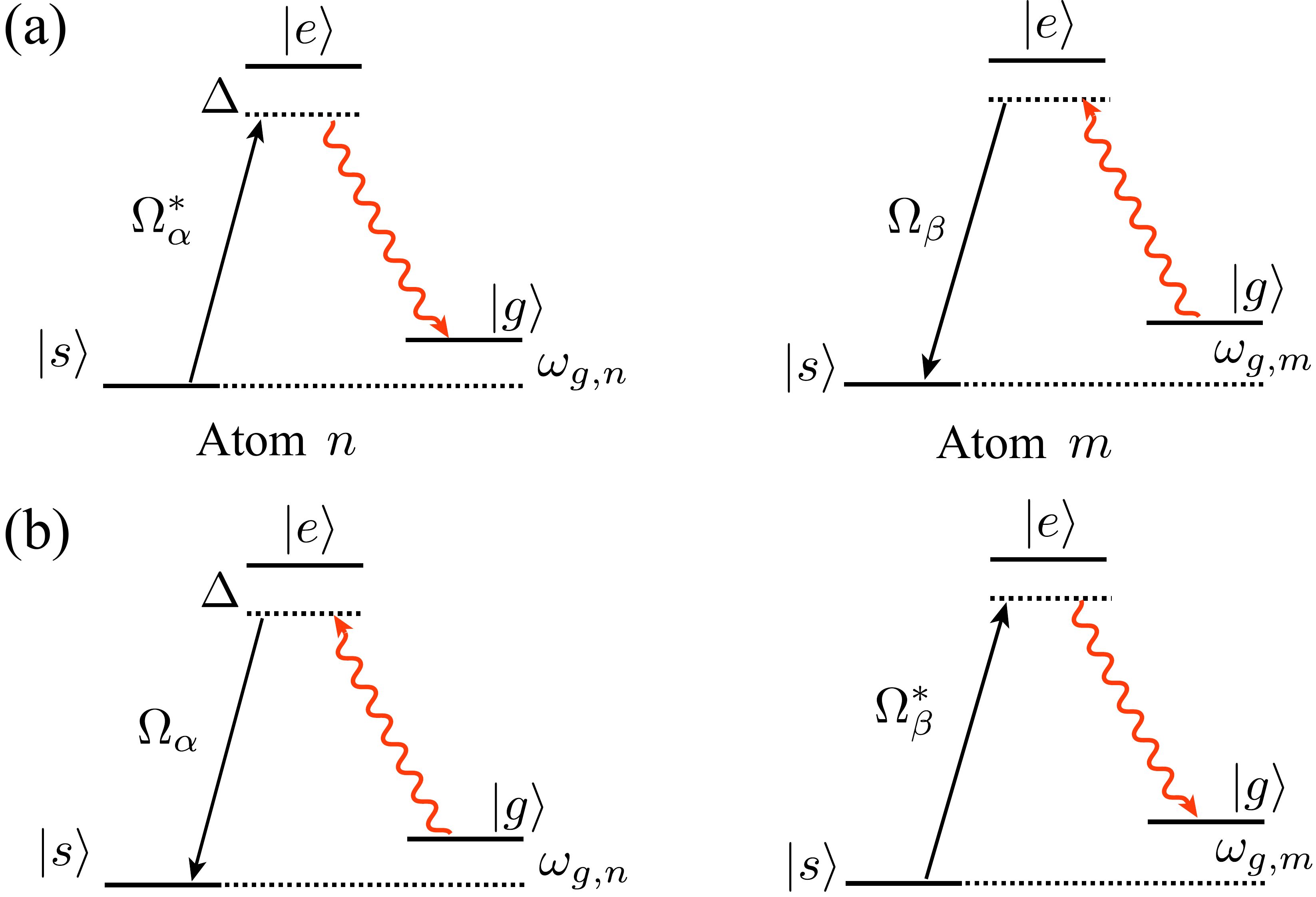}
\caption{Schematics to engineer long-range spin exchange interactions via resonant Raman-scattering processes. Spin exchanges (a) $\ket{s_n, g_m} \rightarrow \ket{g_n, s_m}$ and (b) $\ket{g_n, s_m} \rightarrow \ket{s_n, g_m}$ are allowed only when the condition $\omega_{g,m}-\omega_{g,n}=\tilde{\omega}_{\beta}-\tilde{\omega}_{\alpha}$ is satisfied. $\Omega_\alpha/\Delta$ and $\Omega_\beta/\Delta$ control the exchange rate.}
\label{fig2}
\end{figure}

The time dependence in Eq.~(\ref{eq:XYham}) can be further engineered and simplified. We note that the interaction between two atoms $n$ and $m$ will be highly dependent on the resonant condition $ \omega_{g,m}-\omega_{g,n}=\tilde{\omega}_{\beta}-\tilde{\omega}_{\alpha}$, provided the ground state energy difference $|\omega_{g,n}-\omega_{g,m}|$ is much larger than the characteristic time scale of interactions $|X_\alpha X_\beta^* \tilde{J}|$. The intuitive picture is depicted in Fig.~\ref{fig2}(a): the atom $n$ scatters from sideband $\alpha$ a photon with energy $\omega_\LL+\tilde{\omega}_{\alpha}-\omega_{g,n}$ into the GMs. When this GM photon propagates to the atom $m$, it will only be re-scattered into a sideband $\beta$ that satisfies $\omega_\LL+\tilde{\omega}_{\alpha}-\omega_{g,n}=\omega_\LL+\tilde{\omega}_{\beta}-\omega_{g,m}$, while the rest of the sidebands remain off-resonant. Figure~\ref{fig2} (b) depicts a reversed process. 

For concreteness, we discuss a 1D case where we assume (v) a linear gradient in the ground state energy $\omega_{g,n}\equiv n\delta$, with $\delta$ being the energy difference between adjacent sites. The sidebands will be chosen accordingly such that $\tilde{\omega}_\alpha=\alpha\delta$, with $\alpha\in \Z$.  

Summing up, with all these assumptions (i-v), the resulting effective Hamiltonian Eq.~(\ref{eq:XYham}) can finally be rewritten as follows:
\begin{equation}
\label{eq:Hamt}
H_{XY}(t)=\sum_p H_{XY,p} e^{i p \delta t}\,,
\end{equation}
where $H_{XY, p}$ is the contribution that oscillates with frequency $p\delta$. Written explicitly,
\begin{align}
\label{eq:Hamtk}
 H_{XY,p}=\sum_{m,n\neq m}^N\sum_{\alpha,\beta=0}^{m_P-1}X_{\alpha}X_{\beta}^*\tilde{J} \delta_{n-m,\beta-\alpha-p}\sigma_{gs}^m\sigma_{sg}^n\,.
\end{align}

\noindent In an ideal situation, the gradient per site satisfies $\delta\gg |X_\alpha X^*_\beta \tilde{J}|$ such that the contributions of $H_{XY,p}$ $\forall\,p\neq 0$ vanish. Under these assumptions, we arrive at an effective time-independent Hamiltonian
\begin{align}
 \label{eq:XYham2}
 H_{XY}(t)\approx H_{XY,0}=\sum_{m,n\neq m}^N J_{m,n}\sigma_{gs}^m\sigma_{sg}^n\,,
\end{align}
where couplings $J_{m,n}$ can be tuned by adjusting the amplitudes and phases of the sidebands $X_\alpha$ as they are given by
\begin{align}
\label{eq:Jmn0}
J_{m,n}=\sum_{\alpha,\beta=0}^{m_P-1}X_{\alpha}X_{\beta}^*\tilde{J} \delta_{n-m,\beta-\alpha}\,.
\end{align}
It can be shown that the set of equations defined by Eq.~(\ref{eq:Jmn0}) has at least one solution for any arbitrary choice of $J_{m,n}$, i.e., by choosing $\Omega_0\gg\Omega_{\alpha \neq 0}$ and $J_{m,n}\approx (X_{0}X_{n-m}^* + X^*_{0}X_{m-n})\tilde{J}$. More solutions can be found by directly solving the set of non-linear equations Eq.~(\ref{eq:Jmn0}). 

It is important to highlight that multi-frequency driving also enables the possibility to engineer geometrical phases and, therefore, topological spin models. If the pump field propagation is not perfectly transverse, that is $\kk_\LL\cdot \rr_{m(n)}\neq 0$ ($\kk_\LL$ being the wave vector of the Raman field), the effective Hamiltonian Eq.~(\ref{eq:XYham2}) acquires spatial-dependent, complex spin-exchange coefficients via the phase of $X_{\alpha}X_{\beta}^*$ in Eq.~(\ref{eq:Jmn0}); see later discussions.

Beyond an ideal setting, we now stress a few potential error sources. Firstly, for practical situations, the gradient per site $\delta$ will be a limited resource, making Eq.~(\ref{eq:XYham2}) not an ideal approximation. Careful Floquet analysis on time-dependent Hamiltonian in Eqs.~(\ref{eq:Hamt}-\ref{eq:Hamtk}) is required, to be discussed later. 
Secondly, there is an additional Stark shift on state $\ket{s}$ due to the Raman fields
 \begin{align} \label{eq:starkshift}
 \delta\omega_{s}(t)&=-\sum_{\alpha=0}^{m_P-1}\frac{|\Omega_\alpha|^2}{4\Delta}-\sum_{\alpha>\beta}^{m_P-1} \Re \Big[\frac{\Omega_\alpha \Omega_{\beta}^*}{2\Delta} e^{i (\tilde{\omega}_\alpha-\tilde{\omega}_\beta)t} \Big]\,,
\end{align}
where $\Re[.]$ indicates real part. We note that the time-independent contribution in Eq.~(\ref{eq:starkshift}) can be absorbed into the energy of $\omega_s$ without significant contribution to the dynamics, while the time-dependent terms may be averaged out over the atomic time scales that we are interested in. 
We will present strategies for optimizing the choice of $\delta$, and minimizing detrimental effects due to undesired time-dependent terms in Eqs.~(\ref{eq:Hamt}) and (\ref{eq:starkshift}) in later discussions.

\begin{figure}
\centering
\includegraphics[width=0.8\columnwidth,keepaspectratio]{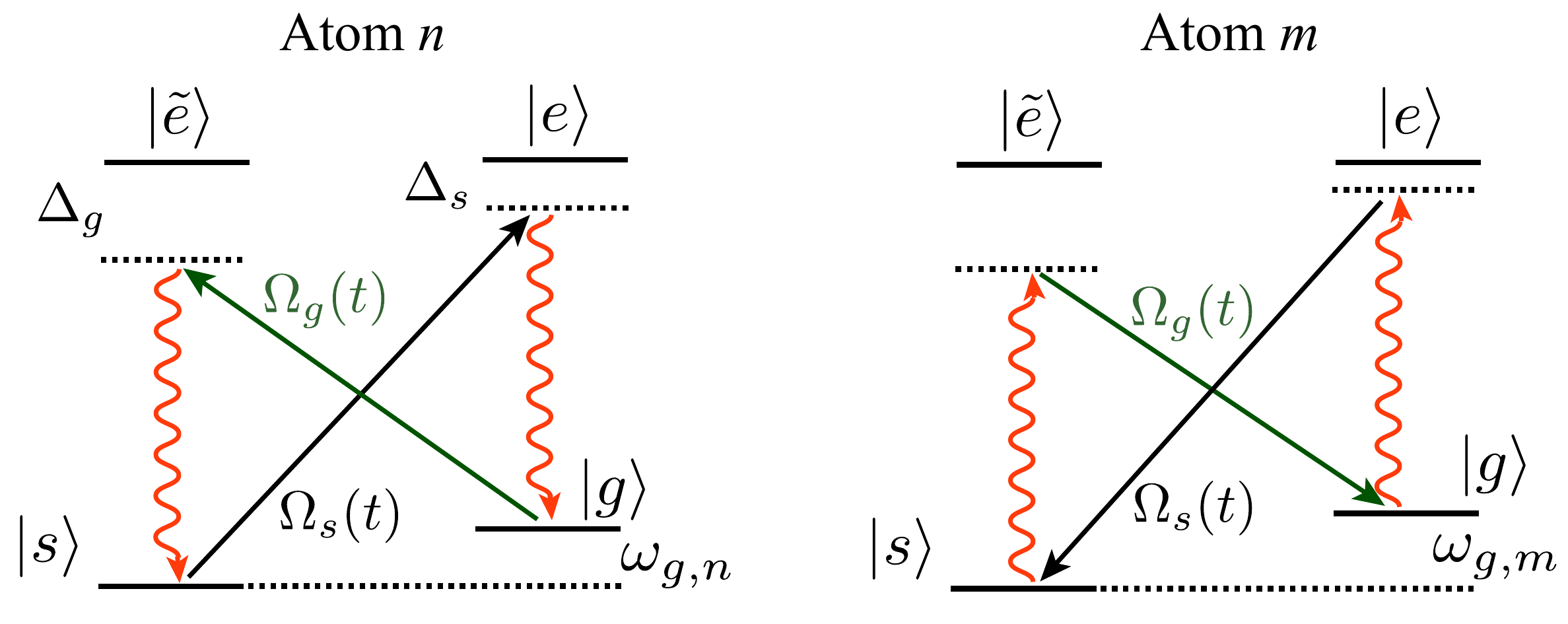}
\caption{Atomic ``butterfly'' level structure. Two pump fields $\Omega_s$ and $\Omega_g$, tuned to couple to the same GM photon, are introduced to control XX and YY interactions independently.}
\label{fig2b}
\end{figure}

\subsection*{Independent control of XX and YY interactions}

So far, we can fully engineer a $XY$ Hamiltonian with equal weight between $XX$ and $YY$ terms. We now show flexible control of $XX$ and $YY$ interactions with slight modifications in the atomic level structure and the Raman-addressing scheme. In particular, we employ a \emph{butterfly}-like level structure where there are two transitions, $\ket{g}\leftrightarrow \ket{e}$ and $\ket{s}\leftrightarrow \ket{\tilde{e}}$, coupled to the same GM, as depicted in Fig.~\ref{fig2b}. We will use two multi-frequency Raman pump fields, $\Omega_g(t)$ and $\Omega_s(t)$, to induce $\ket{g}\leftrightarrow \ket{\tilde{e}} \leftrightarrow \ket{s}$ and $\ket{s}\leftrightarrow \ket{e} \leftrightarrow \ket{g}$ two-photon Raman transitions, respectively. 

For example, to control $XX $ or $YY$ interactions, we require that the two pump fields induce spin flips with equal amplitude, i.e., $\sigma_{gs} \pm  \sigma_{sg}$. This is possible if we choose the main frequencies of the pumps ($\omega_{L,g}$ and $\omega_{L,s}$) such that $\omega_{L,s}=\omega_{L,g} + 2\omega_g$, and match their amplitudes such that $|\Omega_{g,\alpha}|/\Delta_{g} = |\Omega_{s,\alpha}|/\Delta_{s}$, where $\Delta_{s}=\omega_{e}-\omega_{\LL,s}$, $\Delta_{g}=\omega_{\tilde{e}}-(\omega_{\LL,g}+\omega_g)$, and $|\Delta_{s,g}| \gg |\Omega_{s,g}|$. 

Following the procedure to adiabatically eliminate the excited states as well as the GMs, we arrive at the following Hamiltonian (SI Appendix \ref{secSM:derivation}),
\begin{align}
 \label{eq:XYham33}
 H_{XX,YY,0}=\sum_{m,n> m}^N \Big[J_{m,n}(\sigma_{gs}^m+e^{i\phi_{gs}}\sigma_{sg}^m)(\sigma_{sg}^n+e^{-i\phi_{gs}}\sigma_{gs}^n)+\mathrm{h.c.}\Big]\,,
\end{align}
where $\phi_{gs}$ is the relative phase between the pumps fields $\Omega_{g,s}$. Assuming the laser beams that generate the Raman fields are co-propagating or are both illuminating the atoms transversely, i.e., $\kk_\LL \cdot \rr_{m,n} = 0$, we can generate either X or Y components, $(\sigma_{sg}^m \pm  \sigma_{gs}^m)$, by setting the phase $\phi_{gs} = 0$ or $\pi$; more exotic combinations are available with generic choice of $\phi_{gs}$. Moreover, if the laser beams are not co-propagating, they create spatially dependent phases $\phi_{gs,m}$. This can create site dependent $XX$, $YY$ or $XY$ terms.

\begin{figure}
\centering
\includegraphics[width=0.8\columnwidth,keepaspectratio]{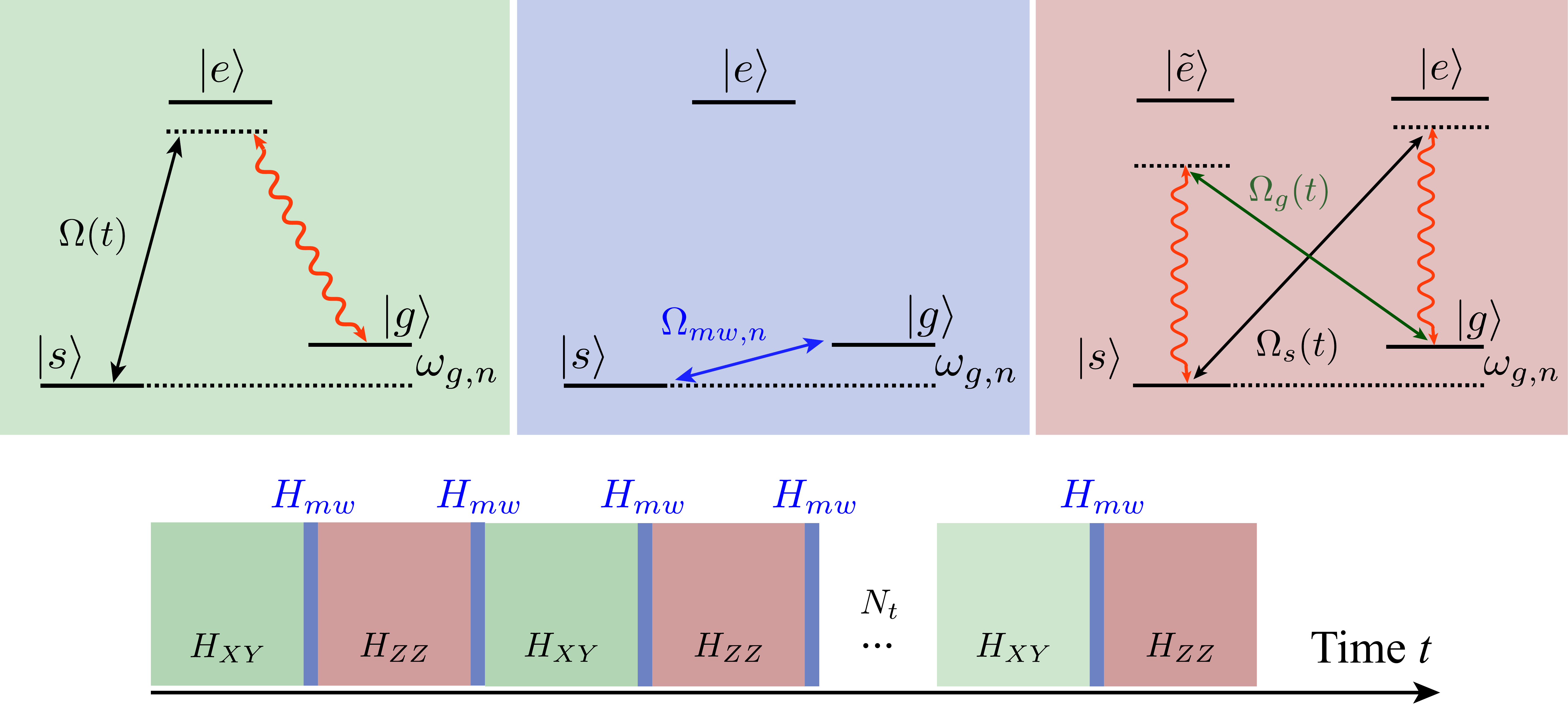}
\caption{Scheme for generating an XXZ spin Hamiltonian using a stroboscopic evolution. The scheme contains periodic applications of a multi-frequency Raman field to induce the $H_{XY}$ interaction (in green), two fast microwave pulses (or optical two-photon transition) forming $H_{mw}$ that uniformly rotate the spin basis  $\{\ket{g}_n,\ket{s}_n\}\leftrightarrow \{(\ket{g}_n+\ket{s}_n)/\sqrt{2},(\ket{g}_n-\ket{s}_n)/\sqrt{2}\}$ back and forth (in blue), and a \emph{buterfly}-like pumping scheme that applies $H_{ZZ}$ in the rotated basis.}
\label{fig3}
\end{figure}

\subsection*{Independent control of ZZ interactions \label{sec:ZZ}}

An independently controlled $ZZ$ Hamiltonian, in combination with arbitrary $XY$ terms, would allow us to engineer a large class of $XXZ$ spin models, i.e., 
\begin{align}
 \label{eq:XXZ}
 H_{XXZ}&=H_{XY}+H_{ZZ}= \sum_{m,n> m}^N \Big[(J_{m,n}^{\mathrm{xy}}\sigma_{gs}^m\sigma_{sg}^n+\mathrm{h.c.})+
 J_{m,n}^{\mathrm{z}}\sigma_z^m\sigma_z^n\Big]
 \,.
\end{align}

In Refs.~\cite{douglas15a,gonzaleztudela15c}, it was shown that $ZZ$ interaction can be created by adding an extra pump field to the $\ket{g}\leftrightarrow\ket{e}$ transition in Fig. \ref{fig1}(c). However, as $ZZ$ terms in this scheme~\cite{douglas15a,gonzaleztudela15c} do not involve flipping atomic states, it is not directly applicable to our multi-frequency pump method. Nonetheless, since we can generate $XX$ and $YY$ interactions independently, a straightforward scheme to engineer $H_{ZZ}$ is to use single qubit rotations to rotate the spin coordinates $X\leftrightarrow Z$ or $Y\leftrightarrow Z$, followed by stroboscopic evolutions \cite{jane03a} to engineer the full spin Hamiltonian. 

One way to engineer $H_{ZZ}$ from pure $XY$ terms is to notice that: $R_{x}(\pm\pi/2)\big(\sigma^n_x \sigma^m_x+\sigma^n_y \sigma^m_y\big) R^\dagger_{x}(\pm\pi/2)=\sigma^n_x \sigma^m_x\mp\sigma^n_z \sigma^m_z$ and $R_{y}(\pm\pi/2)\big(\sigma^n_x \sigma^m_x+\sigma^n_y \sigma^m_y\big) R^\dagger_{y}(\pm\pi/2)=\sigma^n_y \sigma^m_y\pm\sigma^n_z \sigma^m_z$,  where we have used the following notation to characterize rotations along the $\mathbf{n}$ axis, i.e., $R_{\mathbf{n}}(\theta)=e^{i \sigma \cdot\mathbf{n} \theta/2}$. This is particularly useful when both $XY$ and $ZZ$ interactions have the same coupling strengths (e.g., the Haldane-Shastry model). For engineering $ZZ$ terms in a generic spin Hamiltonian, one can apply spin-rotations $R_{y}(\pi/2)$, such that $R_{y}(\pi/2)\sigma_x R^\dagger_{y}(\pi/2)=\sigma_z$, to transform arbitrary $XX$ terms into desired $ZZ$ interactions. Spin-rotation can be realized, e.g., with a collective microwave driving $H_{\mathrm{mw}}=\sum_n \Big(\frac{\Omega_{\mathrm{mw}}}{2}\
\sigma_{sg}^n+\mathrm{h.c.}\Big)$, in which a $\pi/2$-microwave pulse rotates the basis $\{\ket{g}_n,\ket{s}_n\}\rightarrow \{(\ket{g}_n+\ket{s}_n)/\sqrt{2},(-\ket{g}_n+\ket{s}_n)/\sqrt{2}\}$. 

Thus, an $H_{\mathrm{XXZ}}$ Hamiltonian can be simulated using the following stroboscopic evolution: $\{H_{XY},H_{ZZ},H_{XY},H_{ZZ},\dots \}$ in $N_t$ step as schematically depicted in Fig.~\ref{fig3}. The unitary evolution in each step $\delta t=t/N_t$ is given by
\begin{align}
e^{-i(H_{XY}+H_{ZZ})\delta t}\approx e^{-i H_{XY}\delta t}e^{-i H_{ZZ}\delta t}\big(1-\frac{i[ H_{XY},H_{ZZ}]\delta t^2}{2}\big)\,,
\end{align}
where we see that the leading error is $O(\delta t^2)$. When repeating this step $N_t$ times, the leading error in the evolution can be bounded by
\begin{align}
 \label{eq:secondorder}
 E_2\le \frac{||[H_{XY},H_{ZZ}]||t^2}{2 N_t}
\end{align}
for $N_t\gg 1$. It can be shown \cite{lloyd96a} that higher order error terms give smaller error bounds. Because of long-range interactions, the commutator $[H_{XY},H_{ZZ}]$ contains up to $N(N-1)(N-2)$ terms 
that are different from $0$ \cite{footnote7}. Thus the scaling of the error in the limit of $N\gg1$ is approximately given by
\begin{align}
 \label{eq:secondorder2}
 E_2\le \frac{N (R J t)^2}{N_t}\,,
\end{align}
where $J=\mathrm{max}[J_{m,n}]$ is the largest energy scale of the Hamiltonian we want to simulate, and $R$ is the approximate number of atoms coupled through the interaction. For example, if  $J_{m,n}$ is a nearest neighbor interaction, $R=1$. If $J_{m,n}\propto 1/|m-n|^\eta$, then $R\propto \sum_{n=1}^N 1/|n|^\eta$, which typically grows much slower than $N$. Since $E_2\propto 1/N_t$, the Trotter error in $N_t$ steps can in principle be decreased to a given accuracy $\epsilon$ by using enough steps, i.e., $N_t\ge \frac{N (R J t)^2}{ \epsilon}$. 

More complicated stroboscopic evolutions may lead to a more favorable error scaling \cite{berry07a,childs11a,berry09a}, though in real experiments 
there will be a trade-off between minimizing the Trotter error and the fidelity of the individual operations to achieve $H_{XY}$ and $H_{ZZ}$. As this will depend on the particular experimental set-up, we will leave such analysis out of current discussions. For the sake of discussion, we will only consider the simplest kind of stroboscopic evolution that we depicted in Fig. \ref{fig3}.

\section*{Engineering spin Hamiltonians for 1D systems: the Haldane-Shastry $S=1/2$ spin chain \label{sec:ex1d}}

In the first example, we engineer a Haldane-Shastry spin Hamiltonian in one dimension~\cite{haldane88a,shastry88a}
\begin{eqnarray}
H_\mathrm{HS} &=& \sum_{m=1}^{N-1}\sum_{n=1}^{N-m} J_n \big[\frac{1}{2}(\sigma^m_{sg}\sigma^{m+n}_{gs} + \mathrm{h.c.}) +  \sigma^m_z\sigma^{m+n}_z \big]
\label{eq:haldaneshastry}
\end{eqnarray}
where $J_n = J_0 / \sin^2 (n\pi/N)$, $J_0=J\pi^2/N^2$ and $N$ is the number of spins. The interaction strength decays slowly with approximately a $1/r^2$-dependence while satisfying a periodic boundary condition. Such spin Hamiltonian is difficult to realize in most physical setups that interact, e.g., via dipolar interactions.

\begin{figure}
\centering
\includegraphics[width=0.7\columnwidth,keepaspectratio]{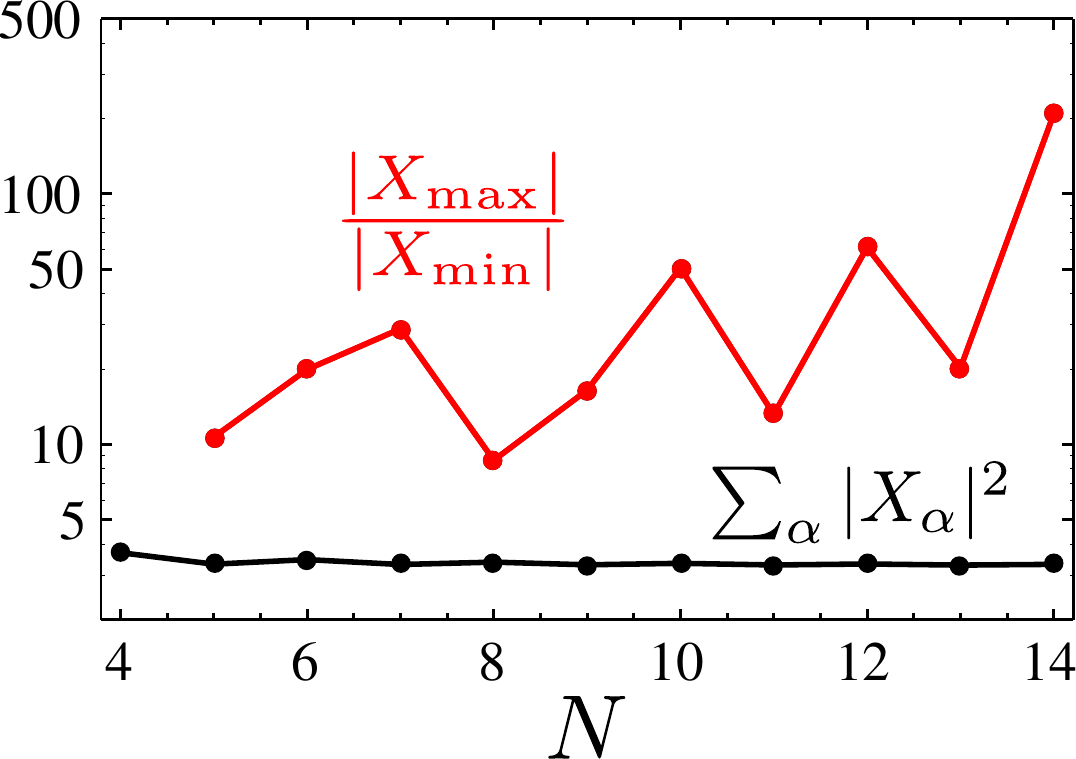}
\caption{Sideband amplitude for Haldane-Shastry model: Total intensity (black) $\sum_\alpha |X_{\alpha}|^2$ and maximum/minimum ratio (red) of sideband amplitudes $|X_\alpha|$ as a function of $N$.}
\label{fig4bis}
\end{figure}

We can engineer the periodic boundary condition and the long-range interaction $J_n$ directly using a linear array of trapped atoms coupled to a PCW. To achieve this, we induce atomic ground state energy shift $m \delta$ according to the spin index $m$, and then uniformly illuminate the trapped atoms with an external pump consisting of $N$ frequency components $\tilde{\omega}_\alpha =  \alpha \delta$, each with an amplitude denoted by $\Omega_\alpha$ and $\alpha=0, 1, ..., N-1$. Regardless of the position of atoms, all pump pairs with frequency difference $n\delta$ contribute to the spin interaction $J_n$. Considering first the XY terms, and according to Eq.~(\ref{eq:XYham2}), we demand
\begin{eqnarray}\label{eq:1dpump}
J_n \approx  
\tilde{J} \sum_{\alpha=0}^{N-n-1} X_\alpha X_{\alpha +n}^*= \frac{J_0}{\sin^2(n \pi/N)},
\end{eqnarray}
where $\tilde{J}$ is the GM photon coupling rate [see Eq.~(\ref{eq:Jmn0})] that we will assume to be a constant for the simplicity of discussions. This requires that the physical size of the spin chain be small compared to the decay length of $\tilde{J}$ (SI Appendix~\ref{secSM:derivation}). That is, $Nd \ll \xi$, where $d$ is the atomic separation. It is then straightforward to find the required pump amplitudes $\Omega_\alpha$ (or equivalently $X_\alpha$) by solving Eq. (\ref{eq:1dpump}) for all $n$. Notice that the system of equations Eq.~(\ref{eq:1dpump}) is overdetermined, and therefore one can find several solutions of it. However, we choose the solution that minimizes the total intensity $\sum_\alpha |\Omega_\alpha|^2$. Figure~\ref{fig4bis} shows that the total intensity converges to a constant value for large $N$, as a result of decreasing sideband amplitudes for decreasing $1/r^2$ interaction strengths. This is 
confirmed in Fig.~\ref{fig4bis} as we see the growth of the ratio between maximum and minimum sideband amplitudes when $N$ increases. The same external pump configuration can also be used to induce the ZZ terms by applying stroboscopic procedures as discussed in the previous section.

\section*{Engineering spin Hamiltonians for 2D systems: topological and frustated Hamiltonians\label{sec:ex2d}}

In the following, we discuss specific examples for engineering 2D spin Hamiltonians that are topologically nontrivial. In particular, we discuss two chiral-flux lattice models that require long-range hopping terms to engineer single particle flat-bands with nonzero Chern numbers, which are key ingredients to realizing fractional quantum hall effects (FQHE) without Landau levels~\cite{neupert11a,sun11a}. Furthermore, our spin-1/2 system interacts like hard core bosons since individual atoms that participate in the spin-exchange process cannot be doubly-excited. With the addition of tunable long-range $ZZ$ interactions, we can readily build a novel many-body system that shall exhibit strongly correlated phases including FQH and supersolid states \cite{wang11a}.

\subsection*{A chiral-flux square lattice model}\label{sec:chiralflux}
The first example can be mapped to a topological flatband model similarly described in Refs.~\cite{neupert11a,sun11a,wang11a}. The topological spin Hamiltonian shall be written as 
\begin{eqnarray}
H_\mathrm{flat} &=& t_1 \sum_{\langle m,n \rangle}  e^{i\phi_{mn}}\sigma_m^\dagger \sigma_n  \pm \,t_2 \sum_{\langle\langle m,n\rangle\rangle} \sigma_m^\dagger \sigma_n +  \mathrm{h.c} (\equiv H_0)\nonumber \\
&+& t_3 \sum_{\langle\langle \langle m,n\rangle\rangle\rangle} \sigma_m^\dagger \sigma_n + \mathrm{h.c} (\equiv H^{'})\,, \label{eq:checkerboard}
\end{eqnarray}
\begin{figure}
\centering
\includegraphics[width=0.85\columnwidth,keepaspectratio]{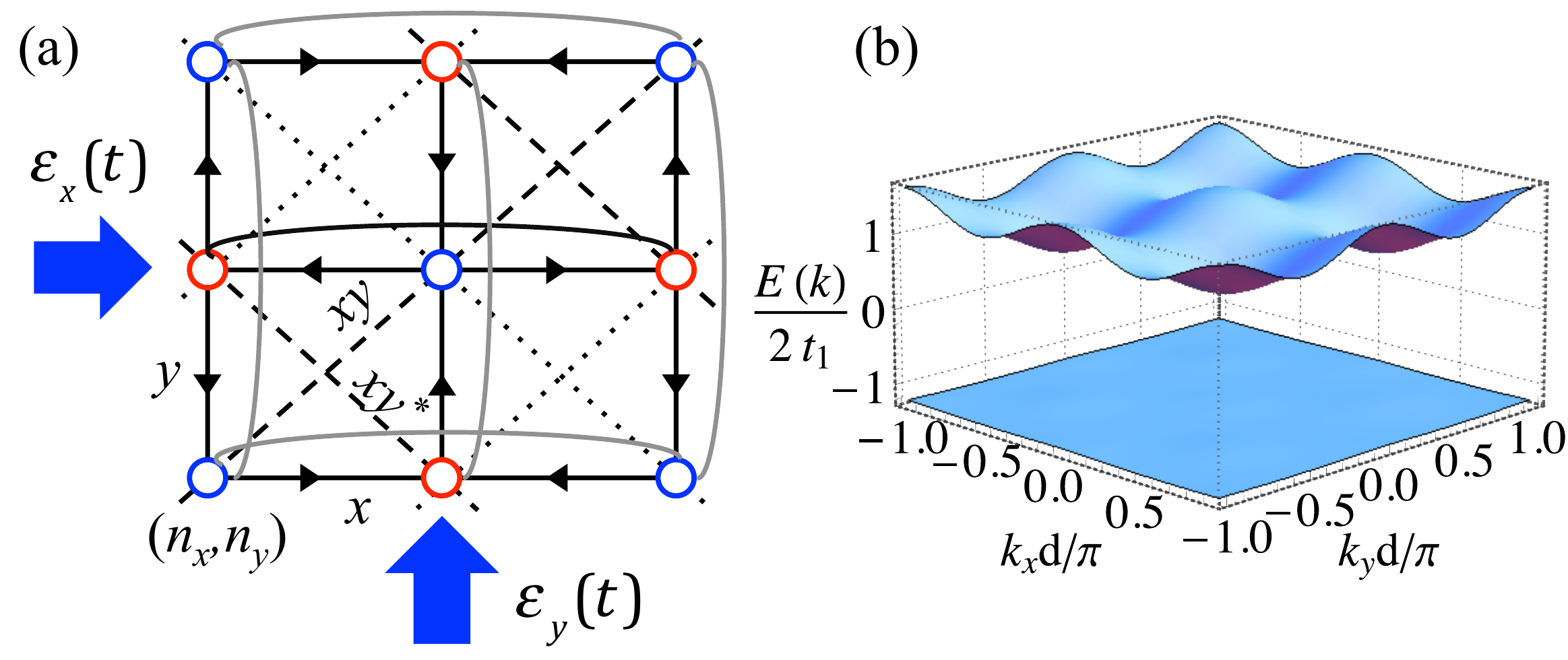}
\caption{Engineering a chiral-flux square lattice.
(a) Two sub-lattices ($n_x+n_y$ odd or even) are marked by blue and red circles, respectively. Solid lines mark the NN hopping with phase gain $\phi$ (arbitrarily tuned) along the direction of the arrows. Dashed (dotted) lines mark the NNN hopping terms (coefficients $\pm t_2$). NNNN long-range hopping along curved lines are included to assist band flattening. Filled arrows indicate the propagation of pump electric fields $\epsilon_y$ and $\epsilon_x$, respectively; see Eqs.~(\ref{eq:exchiral}-\ref{eq:eychiralAmp}). (b) Resulting two-band structure. }
\label{fig4}
\end{figure}

\noindent in which we define $\sigma_m^{\dagger} \equiv \sigma_{sg}^m$ and $\sigma_m \equiv \sigma_{gs}^m$; $\mean{.}$ denotes nearest neighbors (NN) and $t_1$ is the coupling coefficient, $\mean{\mean{.}}$ [$\mean{\mean{\mean{.}}}$] denotes [next-]next-nearest neighours (NNN) [and NNNN respectively] with $t_2$ [$t_3$] being the respective coupling coefficients. The NN coupling phases $\phi_{mn} = \pm \phi$ are staggered across lattice sites, where the phase factor $\phi$ is the one that breaks time-reversal symmetry for $\phi\neq 0,n\pi$ (with $n\in\Z$). Spin exchange between next-nearest neighbors (NNN) has real coefficients $\pm t_2$ with alternating sign along the lattice ``checkerboard'' [see Fig.~\ref{fig4}]. One can show that already $H_0$ has a flatband with non-trivial Chern-number which, choosing $t_2=t_1/\sqrt{2}$, results in a simple band dispersion $E_0(\mathbf{k}) = \pm \sqrt{2} t_1 \sqrt{ 3 + \cos (k_x+k_y) \cos (k_x-k_y) }$. Adding $H^{'}$ to $H_0$ with, e.g.,  $t_3=1/4\sqrt{6} t_1$ allows us 
to engineer an even flatter lower-band whose bandwidth is $\sim 1~\%$ of the band gap. 

We can use an array of atoms trapped within a 2D PCW, as in Fig.~\ref{fig1}(b),  to engineer the Hamiltonian $H_{\mathrm{flat}}$ of Eq.~(\ref{eq:checkerboard}). For simplicity, we assume that there is one atom per site although this is not a fundamental assumption~\footnote{In principle, exact physical separations between trapped atoms do not play a significant role with photon mediated long-range interactions. One may also engineer the spin Hamiltonian based on atoms sparsely trapped along a photonic crystal, even without specific ordering. It is only necessary to map the underlying symmetry and dimensionality of the desired spin Hamiltonian onto the physical system.}. As shown in Fig.~\ref{fig4}(a), we need to engineer spin exchange in four different directions, namely, $\hat{x},\hat{y},\hat{x}\pm\hat{y}$. We first introduce linear Zeeman shifts by properly choosing a magnetic field gradient $\nabla B$ (see SI Appendix \ref{secSM:Zeeman}) such that $\delta_\alpha =  |\mu_B \nabla B \cdot \Delta \mathbf{r}_\alpha|$, where $\mu_B$ is the magnetic moment, $ \Delta\mathbf{r}_\alpha$ are vectors associated with the directions of spin exchange: $\{\Delta\mathbf{r}_\alpha\}_{\alpha=x,y,xy,xy^*} =\{d\hat{x},  d\hat{y}, d(\hat{x}+\hat{y}),d(\hat{x}-\hat{y})\}$, and $d$ is the lattice constant. To activate spin exchange along these directions while suppressing all other processes, we consider a simplest case by applying a strong pump field of amplitude $\Omega_0$ (frequency $\omega_\LL$) to pair with sidebands $|\Omega_\alpha| \ll |\Omega_0|$ of detunings $\tilde{\omega}_\alpha = \delta_\alpha$ to satisfy the resonant conditions. To generate the desired chiral-flux 
lattice, we need to carefully consider the propagation 
phases $\kk_0 \cdot \rr_n$ ($\kk_\alpha \cdot \rr_n$) of the pump field (and sidebands), where $\rr_n = d(n_x\hat{x}+n_y\hat{y})$ 
is the site coordinate and $n_{x, y} \in \mathds{Z}$. In the following we pick $k=k_\alpha=\pi/d$. 

We can generate the required couplings in $H_{\mathrm{flat}}$, term-by-term, as follows:
\begin{itemize}
 \item \emph{Staggered NN coupling along $\Delta\rr_{\alpha=x,y}$}: These are the most important terms as they break time-reversal symmetry. In order to engineer complex coupling along $\Delta\rr_{x}$, we consider the strong pump field to be propagating along $\hat{y}$, that is, $X_0(\rr_n) = \frac{|\Omega_0|}{2\Delta}e^{-i n_y \pi}$. At the NN site $\rr_m = \rr_n+\Delta\rr_x$, it can pair with an auxiliary sideband of detuning $\tilde{\omega}_x= \delta_x = |\mu_B \nabla B \cdot \Delta \rr_x|$ with $X_x (\rr_m) = \frac{|\Omega_1|}{2\Delta}[  e^{-i  n_y \pi} - i \zeta  e^{-i  (n_x+1) \pi}]$. The sideband is formed by two field components in $\epsilon_y(t)$ and $\epsilon_x(t)$, propagating along $\hat{y}$ and $\hat{x}$ respectively [see Fig.~\ref{fig4} and Eqs.~(\ref{eq:eychiral}-\ref{eq:exchiral})], with an amplitude ratio of $\zeta$ and with an initial $\pi/2$ phase difference. These two fields are used to independently control real and imaginary parts of the spin-exchange coefficients. Using Eq.~(\ref{eq:
Jmn0}) under 
the condition that $|\Omega_0| \gg |\Omega_\alpha|$, we see that the NN coupling rate along $\hat{x}$ is
 \begin{equation}
  \label{eq:mm}
  J_{m,n} = \tilde{J} X_0 X_x^* = t_1 \frac{\left[1-i \zeta (-1)^{n_x-n_y} \right]}{\sqrt{1+\zeta^2}} = t_1 e^{\pm i \phi}\,,
 \end{equation}
where $t_1 = \tilde{J}|X_0||X_x|\sqrt{1+\zeta^2}$. This results in the staggered phase pattern with $\phi = \tan^{-1}\zeta$ that can be tuned arbitrarily. Extension to the staggered NN coupling along $\Delta\rr_y$ can be realized by introducing another sideband with detuning $\tilde{\omega}_y= \delta_y = |\mu_B \nabla B \cdot \Delta \rr_y|$ and $X_y(\rr_n + \Delta\rr_y) = -\frac{|\Omega_1|}{2\Delta} [ e^{-i (n_y+1)\pi} + i \zeta e^{-i n_x \pi}]$; see Fig.~\ref{fig4}(a) and Eqs.~(\ref{eq:eychiral}) and (\ref{eq:exchiral}). 
 \item \emph{NNN coupling along $ \Delta\rr_{\alpha =xy, xy^*}$:} The coupling coefficient $\pm t_2$ is real and the sign depends on the sub-lattices. To pair with the pump field $X_0$ at site $\rr_n$, we use two sidebands formed by field components in $\epsilon_x(t)$, propagating along $\hat{x}$ with detunings $\delta_{\alpha=xy, xy^*}$ and $X_\alpha(\rr_n + \Delta \rr_{xy,xy^*}) = \pm \frac{|\Omega_2|}{2\Delta} e^{-i \pi (n_x+1)}$ at NNN sites. The resulting exchange coefficients are $J_{m,n} = \tilde{J} X_0X_{\alpha= xy,xy^*}^* = \mp t_2(-1)^{n_x-n_y}$, forming the required pattern with $t_2 = \tilde{J}|X_0| |X_{xy}|$.
  
\item \emph{NNNN coupling along $2\Delta\rr_{\alpha=x,y}$:} Two sidebands $X_{2x,2y}= |\Omega_3| e^{-i\pi n_y}/2\Delta\,$, propagating along $\hat{y}$ with detunings $2\delta_{\alpha=x, y}\,$, can introduce the real coupling coefficient $t_3 = \tilde{J} |X_0| |X_{2x}|$.  
\end{itemize}

Summing up, all the components in the Raman field can be introduced by merely two pump beams propagating along $\hat{x}$ and $\hat{y}$ directions, respectively. Explicitly writing down the time-dependent electric field $\mathbf{E}(\rr_n, t)$ that generates the desired Raman field $\Omega(t) \equiv \bra{s}\mathbf{d}\cdot\mathbf{E}^*\ket{e}$, we have $\mathbf{E}(\rr_n,t)=\hat{z}\left[\epsilon_x(t) e^{i k \hat{x} \cdot \rr_n} + \epsilon_y(t)e^{i k \hat{y} \cdot \rr_n }\right] e^{ -i \omega_\LL t}$, where $\bra{s}\mathbf{d}\ket{e}$ is the transition dipole moment. For the field propagating along $\hat{y}$, the amplitude reads
\begin{eqnarray}
\epsilon_y(t) = \epsilon_0 + \epsilon_1  (e^{-i \delta_x t} - e^{-i \delta_y t}) + \epsilon_3 (e^{-2 i \delta_{x}t} + e^{-2i \delta_{y} t}). \label{eq:eychiral}
\end{eqnarray}
For the field propagating along $\hat{x}$, we similarly require
\begin{eqnarray}
\epsilon_x(t) = i \zeta \epsilon_1 ( e^{-i \delta_x t} +  e^{-i \delta_y t} )+ \epsilon_2(e^{-i \delta_{xy}t} - e^{-i \delta_{xy*} t}).  \label{eq:exchiral}
\end{eqnarray}
Each term in Eqs.~(\ref{eq:eychiral}-\ref{eq:exchiral}) contributes to specific sideband in the Raman field and $|\Omega_{\alpha}| = \bra{s}\mathbf{d}\cdot\hat{z}\ket{e} \epsilon_{\alpha}$. Pairing individual sidebands to the main Raman field introduced by the leading term $\epsilon_0 \gg \epsilon_{1,2,3}$ leads to the desired spin exchange interactions as discussed above.

We note that there are more ways other than Eqs.~(\ref{eq:eychiral}-\ref{eq:exchiral}) to engineer the spin Hamiltonian. It is also possible to introduce both blue-detuned ($\delta_\alpha>0$) and red-detuned ($\delta_{-\alpha} = -\delta_\alpha$) sidebands in the Raman field to control the same spin-exchange term. That is, 
\begin{equation}
J_{m,n}=\tilde{J} \left[ X_0(\rr_n)X_\alpha^*(\rr_m)+X_0^*(\rr_m)X_{-\alpha}(\rr_n) \right], 
\end{equation}
which has contributions from $X_\alpha$ and $X_{-\alpha}$ of blue- and red-sidebands, respectively.
Arranging both sidebands with equal amplitudes can lead to equal contributions in the engineered coupling coefficient. This corresponds to replacing frequency shifts $e^{- i \delta_\alpha t}$ in Eqs.~(\ref{eq:eychiral}-\ref{eq:exchiral}) with amplitude modulations $\cos \delta_\alpha t$. We may replace the fields by 
\begin{align}
\epsilon_y(t) &= \epsilon_0 + \frac{\epsilon_1}{2}(\cos\delta_x t - \cos\delta_y t) + \frac{\epsilon_3}{2} (\cos2 \delta_{x}t + \cos2 \delta_{y} t), \label{eq:eychiralAmp}\\ 
\epsilon_x(t) &= i \zeta \frac{\epsilon_1}{2} ( \cos \delta_x t +  \cos \delta_y t )+ \frac{\epsilon_2}{2}( \cos\delta_{xy}t - \cos\delta_{xy*} t). \label{eq:exchiralAmp}
\end{align}
In real experiments, amplitude modulation can be achieved by, e.g., the combination of acoustic-optical modulators, and optical IQ-modulators.
\begin{figure}
\centering
\includegraphics[width=0.85\columnwidth,keepaspectratio]{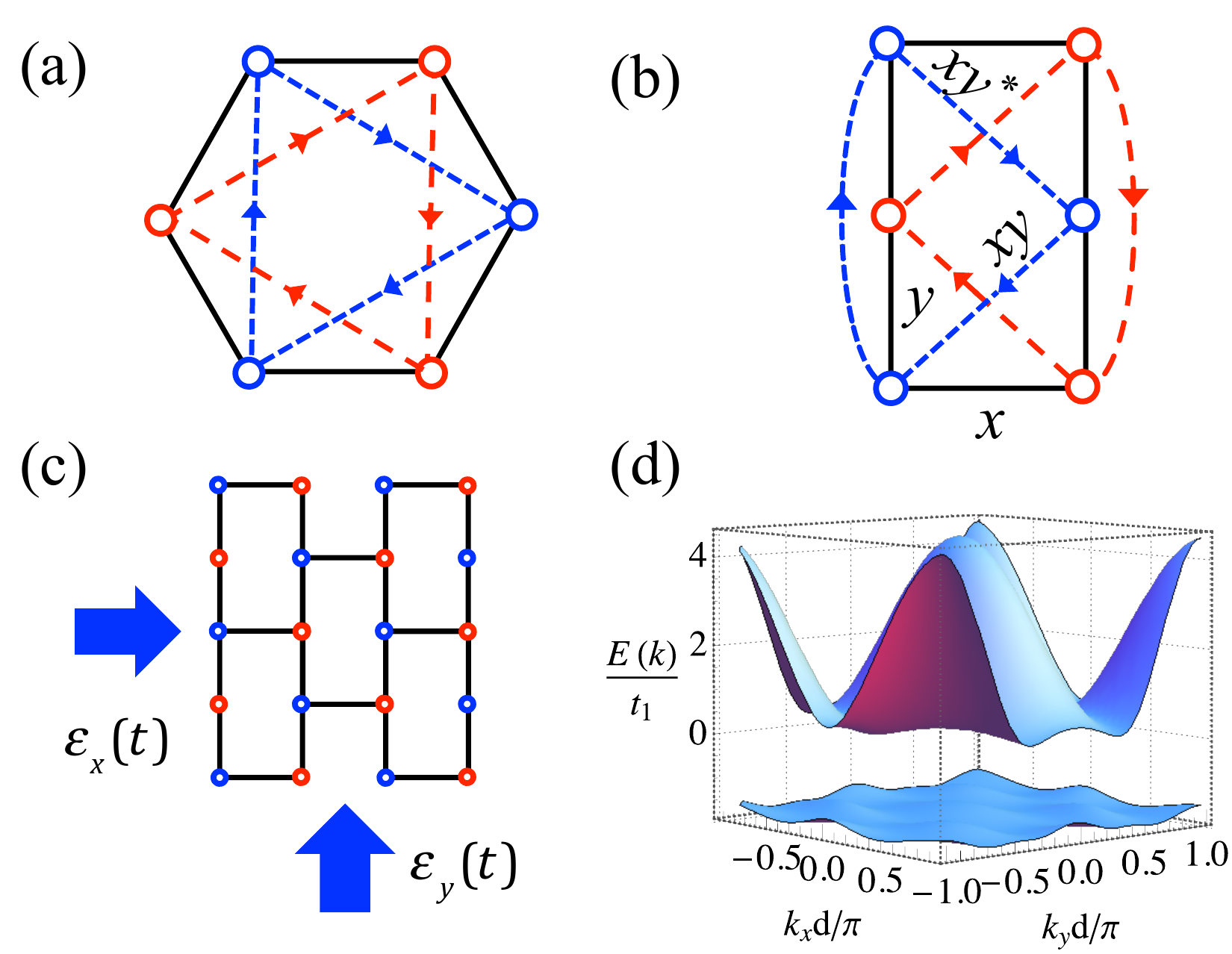}
\caption{Engineering a honeycomb-equivalent topological brick-wall lattice.
(a) Unit cell of a honeycomb lattice. Solid lines mark the NN hopping. Dashed lines mark the NNN hopping with phase gain $\phi$ along the direction of the arrows. (b) Unit cell of a brick-wall lattice. Solid lines indicate the NN hopping as in (a). NNNN hopping (curved dashed lines) and NNN hopping (diagonal dashed lines) corresponds to the complex NNN hopping in (a), making the two models topologically equivalent. (c) Brick-wall lattice. Filled arrows illustrate the pump electric fields; see SI Appendix~\ref{secSM:brickwall}. (d) Band structure of the brick-wall lattice, plotted with $\cos \phi=3\sqrt{3/43}$ \cite{neupert11a}.}
\label{fig5}
\end{figure}
\subsection*{A ``honeycomb''-equivalent topological lattice model}\label{sec:brickwall}
To further demonstrate the flexibility of the proposed platform, we create Haldane's honeycomb model \cite{haldane88b} via a topologically equivalent brick-wall lattice~\cite{tarruell12a,jotzu14a}. 
Here, we engineer the brick-wall configuration using the identical atom-PCW platform discussed in the previous example. Mapping between the two models is illustrated in Figs.~\ref{fig5}(a-b), which contains the following two non-trivial steps: i) generating a \emph{checkerboard-like} NN-exchange pattern in the $\hat{x}$ direction; ii) obtaining NNN (along $\Delta\rr_{xy,xy^*}$) and NNNN (along $2\Delta\rr_{y}$) couplings with the same strength and with a coupling phase $\phi_{mn} = \pm \phi$, which alternates sign across two sub-lattices. Thus, our target Hamiltonian is given by
\begin{eqnarray}
H = t_1\sum_{\langle m,n\rangle} (\sigma_m^\dagger \sigma_n +\mathrm{h.c.} )+ t_2 \sum_{\{m,n\}} ( e^{i\phi_{mn}}\sigma_m^\dagger \sigma_n + \mathrm{h.c.}), \label{eq:Haldane}
\end{eqnarray}
where $\mean{.}$ denotes NN pairs in the brick-wall configuration [see Fig.~\ref{fig5}] and $t_1$ is the coupling coefficient. Note that, for simplicity, we discuss a special case where all NN-coupling coefficients from a brick-wall vertex are identical.
The second summation in Eq.~(\ref{eq:Haldane}) runs over both NNN and NNNN pairs with identical coupling coefficient $t_2$ and alternating phase $\phi_{mn} = \pm \phi$; see Fig.~\ref{fig5}. 

As in the previous case, we use a strong pump field of amplitude $\Omega_0$, propagating along $\hat{y}$, as well as several other sidebands $|\Omega_\alpha|\ll |\Omega_0|$ of detunings $\tilde{\omega}_\alpha=\delta_\alpha$ to generate all necessary spin-exchange terms. Detailed descriptions on engineering individual terms can be found in SI Appendix~\ref{secSM:brickwall}. 

The most important ingredient, discussed here, is that we can generate \emph{checkerboard-like NN coupling} (along $\hat{x}$), with $ J_{m,n} = \tilde{J} X_0 X_x^* = \frac{t_1}{2} \left[1- (-1)^{n_x - n_y} \right]$. This is achieved by using a sideband of detuning $\delta_x$ and amplitude $X_x = \frac{|\Omega|}{4\Delta} [  e^{-i  n_y \pi} + \zeta e^{-i  (n_x+1) \pi} ]$ at position $\rr_m = \rr_n + \Delta \rr_x$, formed by two fields propagating along $\hat{y}$ and $\hat{x}$, respectively. If both fields have the same amplitude ($\zeta =1$), they either add up or cancel completely depending on whether $n_x-n_y$ is odd or even. If one applies the same trick toward \emph{NN coupling along $\hat{y}$}, but with $\zeta \neq 1$, the coupling amplitude modulates spatially in a checkerboard pattern. Essentially all three NN terms around a brick-wall vertex can be independently controlled, opening up further possibilities to engineer, for example, Kitaev's honeycomb lattice model \cite{kitaev2006,feng07}.

For physical implementations, again only two pump beams can introduce all components required in the Raman field, which is very similar to the previous case in either Eqs~(\ref{eq:eychiral}-\ref{eq:exchiral}) or Eqs~(\ref{eq:eychiralAmp}-\ref{eq:exchiralAmp}). Detailed configurations can be found in SI Appendix~\ref{secSM:brickwall}. We stress that by merely changing the way the Raman field is modulated, one can dynamically adjust the engineered spin Hamiltonians and even the topology, as we compare both cases. This is a unique feature enabled by our capability to fully engineer long-range spin interactions.

Moreover, many of the tricks discussed above can also be implemented in 1D PCWs. It is even possible to engineer a topological 1D spin chain, by exploiting long-range interactions to map out non-trivial connection between spins. For example, our method can readily serve as an realistic approach to realize a topological 1D spin chain as recently proposed in Ref.~\cite{grass15}.

\subsection*{XXZ spin Hamiltonian with tunable interaction $1/r^\eta$}\label{sec:Hxxz}

In the last example, we highlight the possibility of engineering a large class of $XXZ$ spin Hamiltonians, which were studied extensively in the literature  \cite{wessel05a,melko05a,boninsegni05a,trefzger08a,buchler07a,hauke10c,maik12a,hauke13a} because of the emergence of frustration related phenomena. A $XXZ$ Hamiltonian is typically written as
\begin{align}
 \label{spin}
H_{XXZ}=&-B \sum_{n} \sigma_n^z \nonumber \\
&+\sum_{n<m} \frac{J}{r_{n,m}^\eta}
\Big[\cos(\theta)\sigma_n^z\sigma_m^z+\sin(\theta)(\sigma_n^x\sigma_m^x+\sigma_n^y\sigma_m^y)\Big]\,,
\end{align}
\noindent where an effective magnetic field $B$ controls the number of excitations, $r_{m,m}=|\rr_{n}-\rr_{m}|$ and the parameter $\theta$ determines the relative strength between the $ZZ$ and $XY$ interactions. This class of spin models has been previously studied, but mostly restricted to nearest neighbours \cite{wessel05a,melko05a,boninsegni05a,trefzger08a} or dipolar ($\eta=3$)
interactions \cite{buchler07a,hauke10c,maik12a}.  

\begin{figure}
\centering
\includegraphics[width=0.8\linewidth]{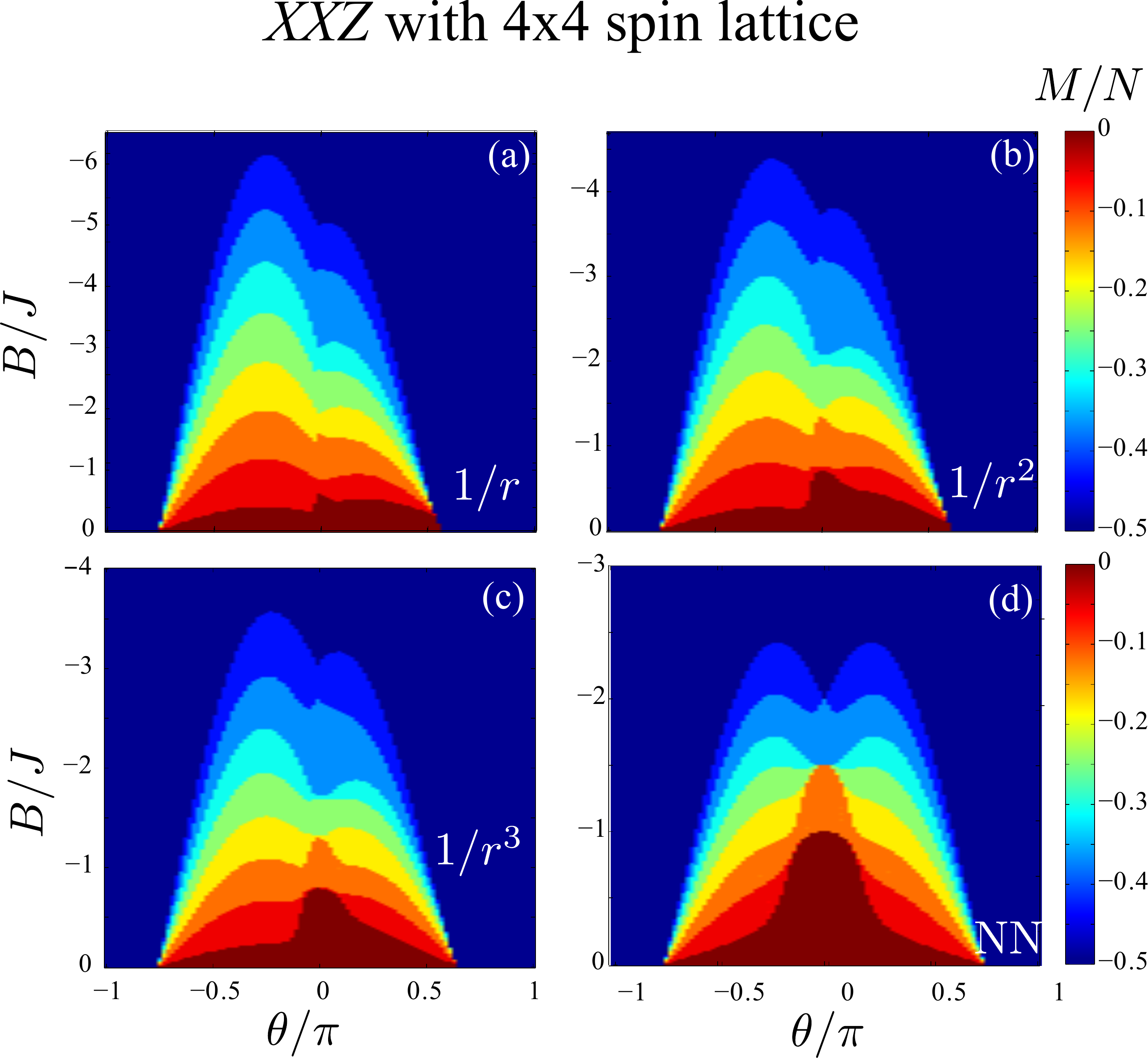}
\caption{Mean magnetization $M/N$ for a system with
$N=16$ atoms in a square lattice, restricted to $N_{\mathrm{exc}}\le8$ excitations and $\eta=1$ (a), $2$ (b), $3$ (c) and NN couplings (d).}
\label{fig6}
\end{figure}

In our setup, one can simulate XXZ models with arbitrary $\eta$ by first introducing unique ground state energy shifts at each of the separation $\rr_{n}-\rr_{m}$ (SI~Appendix~\ref{secSM:Hxxz}), and then applying a strong pump field of amplitude $\Omega_{0}$ together with $N_d$ auxiliary fields $\Omega_\alpha$ of different detunings to introduce spin interactions at each separations \footnote{To simulate a square lattice of $n_s\times n_s$ ($=N$) atomic spins, we find that the number of different distances grows as $N_d=\frac{n_s(n_s+1)-2}{2}$, which is linearly proportional to the number of atoms $N_d\propto N$.}. Moreover, the parameter $\theta$ that determines the ratio between $ZZ$ and $XY$ interaction can be controlled by using different pump intensities in the stroboscopic steps (SI Appendix \ref{secSM:Hxxz}).

To illustrate physics that can emerge in the first experimental setups with only a few atoms, we study the total magnetization of a small square lattice of
$n_s\times n_s$ ($=N$) $16$ atomic spins. We apply exact
diagonalization restricting to $N_{\mathrm{exc}}\le 8$ excitations for $N=16$ spins and cover half of the phase diagram with $B>0$. In Fig. \ref{fig6}, we explore the mean polarization of the system $M/N=\frac{1} {2}\sum_i^{N}\mean{\sigma_i^z}/N$ as a function of $B$ and $\theta$ for $\eta=1$ (a), $2$ (b), $3$ (c) and NN couplings (d). At $\theta=0$, the system behaves classically showing the so-called \emph{devil staircase} \cite{bak82a} of insulating states with different rational filling factors and crystalline structures. With only NN couplings, as shown in Fig.~\ref{fig6} (d), the phase diagram is symmetric with respect to $\theta$. However, long-range hoppings modify the crystal behavior for $\theta\lessgtr0$, making this phase diagram asymmetric; see Figs.~\ref{fig6} (a-c). Longer-range coupling helps the crystal acquire off-diagonal correlations, turning crystalline states into supersolids. This was predicted using dipolar couplings \cite{buchler07a,maik12a}. Nonetheless, experimental observations 
remain elusive \cite{prokof07a}. The higher degree of asymmetry for $\eta=1$-interactions is a strong signature that frustration effects are much more relevant than those with NN or dipolar couplings \cite{shastry81a}. This points at the possibility of observing more stable supersolid phases as well as other frustration-related phenomena such as long-lived metastable states which could serve as quantum memories \cite{trefzger08a}. Another especially interesting arena is the 
behavior of strongly long-range interacting 
systems under non-equilibrium dynamics. It has recently been predicted to yield ``instantaneous'' transmission of correlations after a local quench \cite{hauke13a,richerme14a,jurcevic14}, breaking the so-called Lieb-Robinson bound.

\section*{Limitations and error analysis \label{sec:error}}
Till now, we have mainly focused on how to engineer $H_0$ in an ideal situation. We neglected spontaneous emission or GM photon losses and considered that the energy gradient (or $\delta$, the ground state energy difference between nearest neighboring atoms) can be made very large compared to the interaction energy scales that we want to simulate ($|\delta| \gg |J_{m,m+1}|$). Since the effect of finite cooperativities was considered in detail in Refs. \cite{douglas15a,gonzaleztudela15c}, and their conclusions translate immediately to our extension to multi-frequency pumps, in this work we mainly focus on the effect of finite $\delta$. In addition, we also discuss the effects of AC Stark shifts as in Eq.~(\ref{eq:starkshift}), and its error contributions.

\subsection*{Corrections introduced from higher harmonics: a Floquet analysis \label{sec:floquet}}

We discuss errors and the associated error reduction scheme following a Floquet analysis \cite{goldman14a}, applicable mainly to 1D models. Including all the time-dependent terms in a multi-frequency pumping scheme, we have [Eq.~\ref{eq:Hamt}] $H(t)=\sum_p H_p e^{i p \delta t}$, where $H_p$ represents the part that oscillates at frequency $p\delta$. This Hamiltonian has a period $T=2\pi/\delta$. It can be shown that at integer multiples of $T$, the observed system should behave as if it is evolving under an effective Hamiltonian given by \cite{goldman14a}
\begin{align}
\label{eq:effhamflo}
H_{\mathrm{eff,1}} &\approx  H_0+ \frac{1}{\delta}\sum_p \frac{[H_p,H_{-p}]}{p} \nonumber \\
 &+ \frac{1}{2\delta^2}\sum_p\frac{[[H_p,H_0],H_{-p}]+[[H_{-p},H_0],H_{p}]}{p^2}\,.
\end{align}
This means that the leading error in our simple scheme would be on the order of $J^2/\delta$, where $J$ is the simulated interaction strength. However, we note that if $H_p = \pm H_{-p}$, the leading error term $\sum_p [H_p,H_{-p}]/(p\delta)$ should vanish. In other words, first order error vanishes if $H_p$ is either symmetric or anti-symmetric under a time-reversal operation $\mathcal{T}$. While the original Hamiltonian $H(t)$ doesn't necessarily possess such symmetry, it is possible to introduce a two-step periodic operation $H_\mathrm{2step} = \{ H, \mathcal{T} H, H, \mathcal{T} H , ...\}$ to cancel the first order error while keeping the time-independent part $H_\mathrm{2step,0}=H_0$ identical. This results in 
an effective Hamiltonian in the Floquet picture:
\begin{align}
\label{eq:error}
H_{\mathrm{eff,2}} = H_0+H_{\mathrm{err,2}} \approx H_0+ \frac{4}{\delta^2}\sum_p(-1)^p\frac{[[\tilde{H}_p,H_0],\tilde{H}_{p}]}{p^2}\,,
\end{align}
where $\tilde{H}_p$ is the (operator) Fourier coefficient of the two-step Hamiltonian at frequency $p\delta$ and the leading error reduces to the order of $J^3/\delta^2$. 
  \begin{figure}
  \centering
  \includegraphics[width=0.7\linewidth]{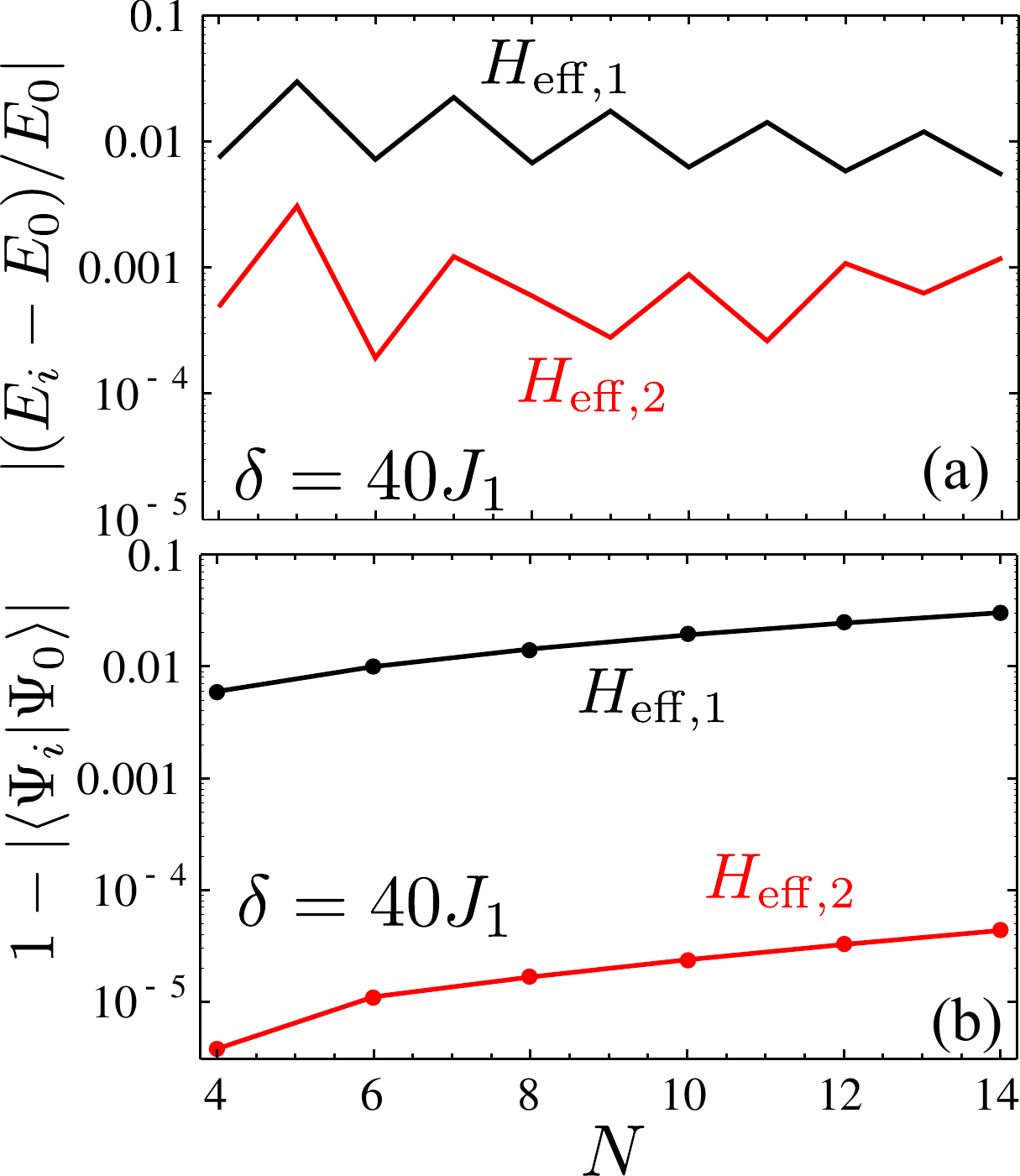}
  \caption{(a) Comparison of ground state energy error $|(E_0-E_i)/E_0|$ and (b) ground state overlap $|\braket{\Psi_0}{\Psi_i}|$ as a function of $N$ for Hamiltonians $H_{\mathrm{eff},1}$ (black) and $H_{\mathrm{eff},2}$ (red) with detuning $\delta/J_1=40$.}
  \label{fig7}
  \end{figure}
To achieve the time-reversal operation, we must reverse the phase of the driving lasers, as well as the sign of the energy offsets between the atoms. Specifically, we can engineer a periodic two-step Hamiltonian by first making the system evolve under presumed $H_0$ (along with other time-dependent terms) for a time interval $T$, and then, for the next time interval $T$, we flip the sign of the energy gradient, followed by reversing the propagation direction of the Raman fields such that $X_\alpha\rightarrow X_{\alpha}^*$ in Eq.~(\ref{eq:Hamt}). As a result, all the time-dependent Hamiltonians $H_{p}$, $\forall p \neq 0$, become $H_{-p}$ in the second step, resulting in $\tilde{H}_p = (-1)^p \tilde{H}_{-p}$ required for error reduction. Whereas, the time-independent Hamiltonian $H_0$ remains identical in the two-step Hamiltonian. See SI Appendix~\ref{secSM:two-step} for more discussions.

\subsection*{Numerical analysis on the Haldane-Shastry spin chain} \label{sec:floquetNumerical}
We now analyze numerically and discuss error on one particular example. For numerical simplicity, we choose the Haldane-Shastry model as its one-dimensional character makes it numerically more accessible. However, the conclusions regarding the estimation of errors can be mostly extended to other models.
 \begin{figure}[t]
  \centering
  \includegraphics[width=0.7\linewidth]{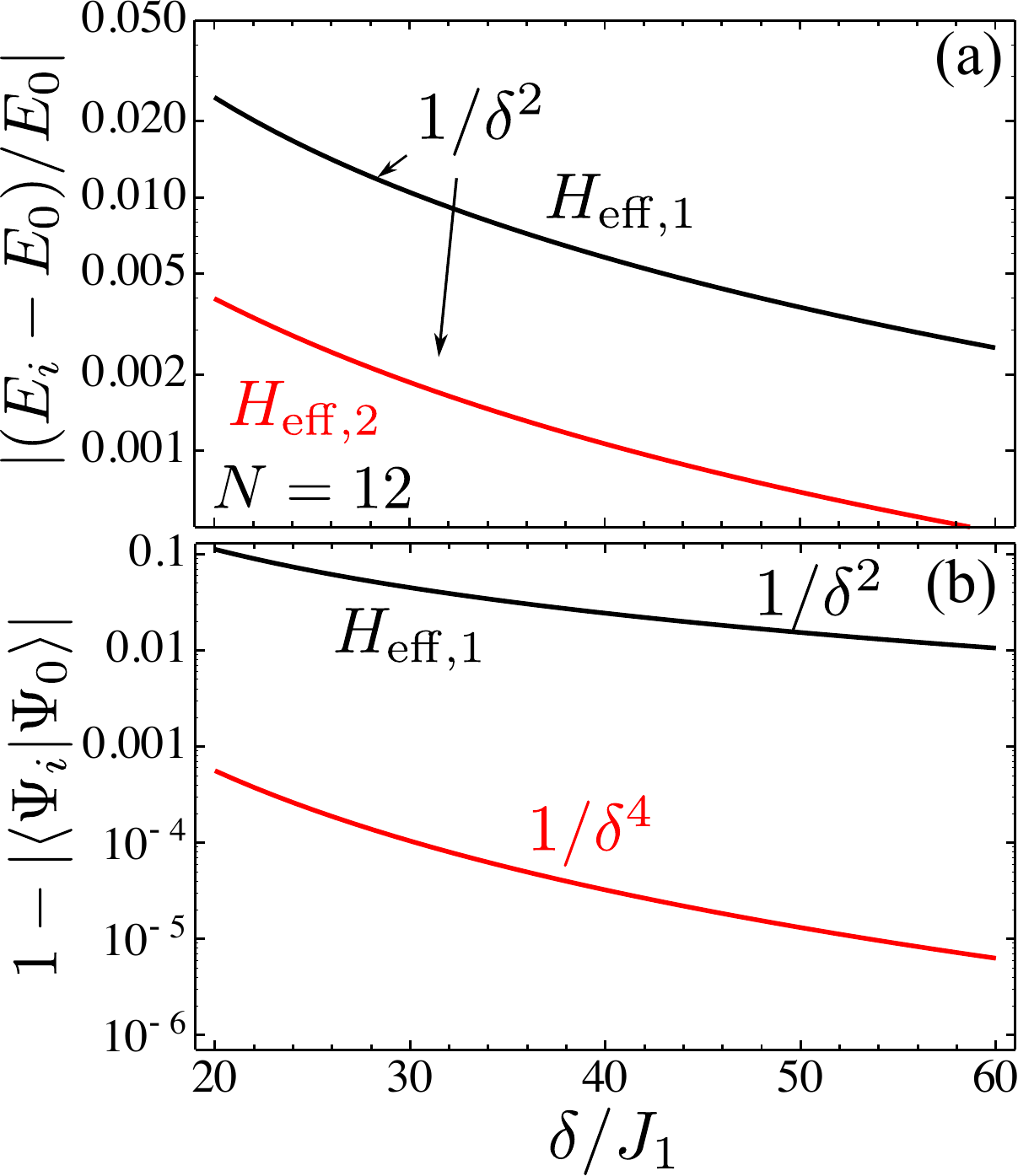}
  \caption{(a) Comparison of ground state energy error $|(E_0-E_i)/E_0|$ and (b) Ground state overlap $|\braket{\Psi_0}{\Psi_i}|$ as a function of $\delta/ J_1$ for Hamiltonians $H_{\mathrm{eff},1}$ (black) and $H_{\mathrm{eff},2}$ (red) for $N=12$ atoms.}
  \label{fig8}
  \end{figure}
As we have shown in Eq.~(\ref{eq:haldaneshastry}), the Haldane-Shastry Hamiltonian is composed by an XY term plus a ZZ term that we can simulate stroboscopically. As we already analyzed the Trotter error due to the stroboscopic evolution, here we focus on the XY part of the Hamiltonian, which reads
\begin{equation}
 H_{\mathrm{HS},\mathrm{xy}}=\sum_{m=1}^N\sum_{n=1}^{N-m} \frac{J_0}{\sin^2(n\pi/N)}\big(\sigma_{sg}^{m}\sigma_{gs}^{m+n}+\sigma_{sg}^{m+n}\sigma_{gs}^{m} \big)\,,
\end{equation}
where $J_0=J\pi^2/N^2$. Following the prescribed engineering steps, the total time dependent Hamiltonian resulting from multiple sidebands can be written as $H(t)=\sum_p H_p e^{i p \delta t}$, with 
\begin{align}
 H_p=\sum_{m=1}^N\sum_{n=1}^{N-m} \big(J_{n,(p)}\sigma_{sg}^{m}\sigma_{gs}^{m+n} +J^*_{n,(-p)}\sigma_{sg}^{m+n}\sigma_{gs}^{m}\big)\,,\label{eq:ht2}
\end{align}
and we have defined $J_{n,(p)}=\sum_{\alpha,\beta=0}^{N-1} X_{\alpha}X_{\beta}^*\delta_{n-p,\beta-\alpha}$.Here, $X_{\alpha}$ are fixed such that $H_0=H_{\mathrm{HS},\mathrm{xy}}$.
  
To illustrate the effect of error cancellations, we consider first a scenario where we directly apply Eqs.~(\ref{eq:ht2}). To leading order, the effective Hamiltonian is $H_\mathrm{eff,1}$ as in Eq.~(\ref{eq:effhamflo}).
We then analyze the two-step driving, using the effective Hamiltonian $H_\mathrm{eff,2}$ in Eq.~(\ref{eq:error}), with $\tilde{H}_p$ given by SI Appendix Eqs.~(\ref{eqSM:even})-(\ref{eqSM:odd}). 

We calculate the ground state energies and eigenvectors of $H_0$,  $H_{\mathrm{eff},1}$ and $H_{\mathrm{eff},2}$, that we denote as $E_{0,1,2}$ and $\ket{\Psi_{0,1,2}}$ respectively, for different number of atoms and different ratios of $\delta/J_0$. The results are shown in Figs. \ref{fig7} and \ref{fig8}. In panels (a) we show the error in absolute value with respect to the ideal Hamiltonian $H_0$. Interestingly, due to particular structure of $\ket{\Psi_0}$ and $H_p$, one can show that $\bra{\Psi_0}[H_p,H_{-p}]\ket{\Psi_0}\approx 0$ and the first order correction to the energy vanishes. This is confirmed in Fig. \ref{fig8}~(a), where we found that the error actually scales with $1/\delta^2$.

Moreover, it is also enlightening to compare the overlap of the ground states as shown in Figs.~\ref{fig7}(b)-\ref{fig8}(b). We only compute the even atom number configuration as the odd ones are degenerate and therefore the ground state is not uniquely defined. We see that the ground state overlap of $H_{\mathrm{eff},2}$ is several orders of magnitude better than the one with $H_{\mathrm{eff},1}$. Moreover, its dependence on $\delta$ is better than the $1/\delta^2$ expectation.

\subsection*{The role of time-dependent Stark shifts in the error analysis \label{sec:stark}}

In the previous discussions,we have dropped the contribution of the time-dependent Stark shifts
\begin{align}
\label{eq:acshift}
H_{\mathrm{ac}}(t)=-\sum_n\sum_{\alpha>\beta}^{m_P-1} \Re[\frac{\Omega_\alpha\Omega^*_\beta}{2\Delta} e^{i\tilde{\omega}_{\alpha,\beta}t} ]\sigma_{ss}^n\,,
\end{align}
where $\tilde{\omega}_{\alpha,\beta}=\tilde{\omega}_\alpha-\tilde{\omega}_\beta$. In the following, we discuss its role in the effective Hamiltonian, using the Floquet error analysis \cite{footnote5}. The Fourier coefficients of the Stark shifts can be written as
\begin{align}
H_{\mathrm{ac},p} = \sum_n A_p^n \sigma_{ss}^n,
\end{align}
where the on-site amplitude
\begin{align}\label{eq:apn}
A_p^n = - \Delta \sum_{\alpha \neq \beta} X_\alpha X_\beta^* \delta(\tilde{\omega}_{\alpha,\beta} - p \delta) \,;
\end{align}
 $A_p^n$ may be site-dependent if the phase differences between the Raman fields $\Omega_\alpha$ vary across sites. Comparing Eq.~(\ref{eq:apn}) with Eq.~(\ref{eq:Jmn0}), we see that $A^n_p \sim O(J\Delta/\tilde{J})$ may be even larger than the engineered interaction  $J \equiv \mathrm{max}[J_{m,n}]$ if $\Delta \gtrsim \tilde{J}$. 
In the two-step driving scheme, we replace the amplitude $A_p^n$ by $\tilde{A}_p^n$ according to SI Appendix~Eqs.~(\ref{eqSM:even}) and (\ref{eqSM:odd}).

As discussed in the previous section, with two-step driving, the leading Stark-shift error contribution only appears in the second order
\begin{align}
\label{eq:acerror}
H_{\mathrm{err,2}} \approx \frac{4}{\delta^2}\sum_p(-1)^p\frac{[[\tilde{H}_p + \tilde{H}_{\mathrm{ac},p},H_0],\tilde{H}_{p} +\tilde{H}_{\mathrm{ac},p}]}{p^2},
\end{align}
where $\tilde{H}_{\mathrm{ac},p}$ are the Fourier coefficients of the two-step Stark-shift Hamiltonian. Only the following nested commutators, $[[\tilde{H}_{\mathrm{ac},p},H_0], \tilde{H}_{\mathrm{ac},p}]$, $[[ \tilde{H}_{\mathrm{ac},p},H_0],\tilde{H}_{p}]$ and $[[\tilde{H}_{p},H_0], \tilde{H}_{\mathrm{ac},p}]$ are related to the time-dependent Stark shifts, and should be evaluated in various configurations: 

\emph{Generic Hamiltonians with translational invariance.} By translational invariance, we mean that there are no site-dependent spin interactions, and the spin-exchange coefficients remain identical as we offset the spin index by one or more. This means that all components in the pump field should drive the system with uniform optical phases as in the Haldane-Shastry model discussed above. The resulting Fourier coefficients $\tilde{H}_{\mathrm{ac},p}  = \tilde{A}_p  \sum_n \sigma^{n}_{ss}$ would have identical effect on all spins ($\tilde{A}_p$ being a constant amplitude). As a result, the above mentioned commutators vanish, suggesting that the error by $\tilde{H}_{ac}(t)$ averages out to zero in the Floquet picture. In the \emph{butterfly} scheme, however, both $\ket{g}$ and 
$\ket{s}$ states are pumped and they may be shifted differently. This leads to slight modifications in the engineered XX and YY terms; see SI Appendix \ref{secSM:error}. 

\emph{Models containing sub-lattices.} For topological models that contain sub-lattices, as in our examples, the pump fields are not perfectly transverse and Stark shifts are site-dependent, resulting in non-vanishing error. However, we also note that the coupling coefficients appearing in the commutators $[[\tilde{H}_{\mathrm{ac},p},H_0], \tilde{H}_{\mathrm{ac},p}]$ and $[[\tilde{H}_{p},H_0], \tilde{H}_{\mathrm{ac},p}] ([[\tilde{H}_{\mathrm{ac},p},H_0], \tilde{H}_{p}]) $ are on the order of $ J|\Omega|^4/\Delta^2$ and $J^2 |\Omega|^2/\Delta$, respectively. For realistic PCW realizations, we may set $\tilde{J}/\Delta \gtrsim O(1)$.
Since $A_p^n/J_{m,n} \sim O(\Delta / \tilde{J})$, the energy scales of the commutators are all on the order $\lesssim J^3$, and the energy scale in $H_\mathrm{err,2}$ will be $\lesssim J^3/\delta^2$. Therefore, for $ \delta \gg J$, the Stark-shift terms may be ignored.

\emph{Stark shift dominated regime.} It may be possible that our sub-lattice models be purposely driven with large amplitude pumps such that $|\Omega|^2/\Delta \gtrsim \delta$. Stark shift contributions would become important in the resulting spin dynamics. However, if we choose a large pump detuning $\Delta > \tilde{J}$, the dominant error contribution (recall that $J\sim \frac{\tilde{J}}{\Delta}\frac{|\Omega|^2}{\Delta}$) comes from the $[[\tilde{H}_{\mathrm{ac},p},H_0], \tilde{H}_{\mathrm{ac},p}]$ term, which can be written in the following simple form
\begin{align}
\label{eq:mainacerror}
H_\mathrm{err,2} &\approx  \frac{4}{\delta^2}\sum_p \frac{(-1)^p}{p^2}[[\tilde{H}_{\mathrm{ac},p},H_0], \tilde{H}_{\mathrm{ac},p}] \\ \nonumber
&\approx \sum_{m,n} \tilde{A}_{m,n} (J_{m,n} \sigma_{gs}^m\sigma^n_{sg} + h.c.).
\end{align}
Here, the site-dependent Fourier coefficient is defined as $\tilde{H}_{\mathrm{ac},p} \equiv \sum_n \tilde{A}^n_p \sigma^{n}_{ss}$ and $\tilde{A}_{m,n}=\sum_p \frac{4(-1)^{p+1}}{\delta^2 p^2}(\tilde{A}^m_p - \tilde{A}^n_p)^2 \neq 0$ in general for interactions $J_{m,n}$ that are not translationally invariant. In a special case that only two sub-lattices are present, as in our examples, we note that $\tilde{A}_{m,n}$ may only depend on the distance $r_{m,n}$ and is site-independent. This `error' term would then \emph{uniformly} modify the XY coupling strengths to a new value
\begin{align}
\label{eq:newJmn}
J_{m,n}^{' } = (1+\tilde{A}_{m,n})J_{m,n}\,.
\end{align}
The next leading order errors are due to $[[\tilde{H}_{p},H_0], \tilde{H}_{\mathrm{ac},p}]/\delta^2$ and $[[\tilde{H}_{\mathrm{ac},p},H_0], \tilde{H}_{p}]/\delta^2$ terms, which are on the order of $J^2 |\Omega|^2/(\Delta\delta^2)$ and are a factor of $\sim \tilde{J}/\Delta$ smaller than the leading Stark shift contribution. This suggests we can always increase the detuning $\Delta$, while keeping $|\Omega|/\Delta$ constant, to reduce the error  contribution. For more discussions, see SI Appendix~\ref{secSM:error}. 

\section*{Conclusions \& Outlook \label{sec:conclusion}}

In this manuscript, we have shown that atom-nanophotonic systems present appealing platforms to engineer many-body quantum matter by using low-dimensional photons to mediate interaction between distant atom pairs. We have shown that, by introducing energy gradients in one and two-dimensions and by applying multi-frequency Raman addressing beams, it is possible to engineer a large class of many-body Hamiltonians. In particular, by carefully arranging the propagation phases of Raman beams, it is possible to introduce geometric phases into the spin system, thereby realizing nontrivial topological models with long-range spin-spin interactions.

Another appealing feature of our platform is the possibility of engineering periodic boundary conditions, as explicitly shown in the 1D Haldane-Shastry model, or other global lattice topology by introducing long-range interactions between spins located at the boundaries of a finite system. Using 2D PCWs, for example, it is possible to create novel spin-lattice geometries such as M\"{o}bius strip, torus, or lattice models with singular curvatures such as conic geometries \cite{biswas14a} that may lead to localized topological states with potential applications in quantum computations.

We emphasize that all the pairwise-tunable interactions can be dynamically tuned via, e.g., electro-optical modulators at time scales much faster than that of characteristic spin interactions. Therefore, the spin interactions can either be adiabatically adjusted to transform between spin models or even be suddenly quenched down to zero by removing all or part of the Raman coupling beams. We may monitor spin dynamics with great detail: after we initially prepare the atomic spins in a known state by, say, individual or collective microwave addressing, we can set the system to evolve under a designated spin Hamiltonian, followed by removing all the interactions to ``freeze'' the dynamics for atomic state detection. Potentially, this allows for detailed studies on quantum dynamics of long-range, strongly interacting spin systems that are driven out-of-equilibrium. The dynamics may be even richer since the spins are weakly coupled to a structured environment via photon dissipations. We expect such a new platform 
may bring novel opportunities to the study of quantum thermalization in long-range many-body systems, or for further understanding of information propagation in a long-range quantum network.

\section*{Acknowledgements}
We gratefully acknowledge discussions with T. Shi and Y. Wu. The work of CLH and HJK was funded by the IQIM, an NSF Physics Frontier Center with support of the Moore Foundation, by the AFOSR QuMPASS MURI, by the DoD NSSEFF program, and by NSF PHY1205729. AGT and JIC acknowledge funding by the EU integrated project SIQS. AGT also acknowledges support from Alexander Von Humboldt Foundation and Intra-European Marie-Curie Fellowship NanoQuIS (625955).

\bibliographystyle{pnas2011} 
\bibliography{Sci_hjk,books}

\begin{thebibliography}{10}
\bibitem{footnote7}
More precisely, $[H_{XY},H_{ZZ}] = -2\sum_{m\neq n}^N J_{m,n}^{xy}
  \Delta^z_{m,n}\sigma_{gs}^m\sigma_{sg}^n$, where $\Delta^z_{m,n} \equiv
  \sum_{q \neq m,n} (J_{q,m}^z-J_{q,n}^z) \sigma_{z}^q$. Assuming a highly
  frustrated state, we simply estimate that $| \mean{\Delta^z_{m,n}}| \lesssim
  JR $ and $||[H_{XY},H_{ZZ}]|| \leq 2(JR)^2$.
\bibitem{footnote5}
Here we discuss a special case where only $\ket{s}$ state is shifted. For a
  general case when both $\ket{g}$ and $\ket{s}$ states are shifted, see SI
  Appendix~\ref{secSM:error}.
\end{thebibliography}

\subsection*{Supporting Information (SI)}
\appendix
In Appendix~\ref{secSM:derivation}, we give a detailed derivation of the time-dependent effective Hamiltonian of Eq. (\ref{eq:XYham}) of the main text. In Appendix~\ref{secSM:Zeeman}, we discuss how to properly introduce ground state energy shifts to engineer generic spin models in two-dimensions. In Appendix~\ref{secSM:brickwall} and~\ref{secSM:Hxxz}, we describe in detail the PCW and the pump field configurations to engineer a topological brick-wall lattice model and a XXZ model with $1/\rr^\eta$-dependence, respectively. In Appendix~\ref{secSM:two-step}, we discuss in detail how to engineer a two-step Hamiltonian. Last, in Appendix~\ref{secSM:error}, we additionally discuss the role of AC Stark shift in a \emph{butterfly} scheme.

\section{Complete derivation of final time-dependent Hamiltonian\label{secSM:derivation}}

The PCWs support localized one or two-dimensional photonic Guided Modes (GMs), which can be described by a Hamiltonian (using $\hbar=1$): 
\begin{equation}
 H_{\mathrm{GM}}=\sum_{\kk}\omega_\kk a^\dagger_\kk a_\kk\,
\end{equation}
where $\omega_\kk$ is the dispersion relation of the GMs.  Neglecting counter-rotating terms, the light-matter Hamiltonian can be written as follows:
\begin{equation}
 H_{\mathrm{lm}}=\sum_{\kk,n}g_\kk (\rr_n)a_\kk \sigma_{eg}^n+\mathrm{h.c.}\,, 
\end{equation}
where $g_\kk (\rr_n)=g_\kk e^{i \kk\cdot \rr_n}$ is the single-photon coupling constant. 
The atomic Hamiltonian is given by
\begin{equation}
H_{\aa}=\sum_n (\omega_e \sigma_{ee}^n+ \omega_{g,n}\sigma_{gg}^n)\,,
\end{equation}
where it is important to highlight that we introduce a site-dependent energy in the hyperfine level $\ket{g}_n$ that can be achieved, e.g., by introducing a magnetic field gradient (or a Stark-shift gradient) in either one or two dimensions as depicted in Fig.\ref{fig1}(a-b) in the main text. This site-dependent energy, together with a multi-frequency driving for $\ket{s}_n\leftrightarrow\ket{e}_n$ are the key ingredients of our proposal. Multi-frequency driving with $m_P$ different components can be described through a Hamiltonian:
\begin{equation}
 H_{\mathrm{d}}(t)=\sum_n\big(\frac{\Omega(t)}{2}\sigma_{se}^n e^{i\omega_{L} t}+\mathrm{h.c.}\big)\,,\label{eqSM:driving}
\end{equation}
where we have used the notation $\sigma_{ab}^n=\ket{a}_n\bra{b}$, and $\omega_L$ is the main driving frequency. All components of the driving field are embedded in $\Omega(t)$, that can be written as
\begin{equation}
 \label{eqSM:muli}
 \Omega(t)\equiv  \sum_{\alpha=0}^{m_P-1}\Omega_\alpha e^{i\tilde{\omega}_\alpha t}\,,
\end{equation}
where $\tilde{\omega}_\alpha$ are $m_P$ different frequency detunings ($\tilde{\omega}_{\alpha=0} =0$) and $\Omega_\alpha$ the Rabi frequency that will be used to achieve full control of the atom-atom interactions. The dynamics of the system is described by the sum of all the above Hamiltonians: $H=H_{\aa}+H_{\mathrm{GM}}+H_d(t)+ H_{\mathrm{lm}}$. 

We are interested in the conditions where $|\Delta|=|\omega_e-\omega_L|\gg\Omega$, such that the excited states are only virtually populated. To adiabatically eliminate states $\ket{e}_n$, it is convenient to work in a rotating frame defined by the transformation $U=\exp\big(i( \sum_n\omega_\mathrm{L}\sigma_{ee}^n t+i\sum_{\kk}\omega_\kk \ud{a}_\kk a_\kk )t\big)$, which transforms the Hamiltonian by $H\rightarrow U H U^\dagger - i U\partial_t U^\dagger$. Writing each of the transformed Hamiltonians, we have
\begin{align}
 H_{\mathrm{lm}} &\rightarrow \tilde{H}_{\mathrm{lm}}=\sum_{\kk,n}g_\kk (\rr_n)a_\kk \sigma_{eg}^n e^{i (\omega_\mathrm{L}-\omega_\kk) t}+\mathrm{h.c.}\,,\nonumber \\
 H_{\mathrm{d}} & \rightarrow \tilde{H}_{\mathrm{d}}=\sum_n\big(\frac{\Omega(t)}{2}\sigma_{se}^n +\mathrm{h.c.}\big)\,, \nonumber \\
 H_{\aa} & \rightarrow \tilde{H}_{\aa}=\sum_n (\Delta \sigma_{ee}^n+ \omega_{g,n}\sigma_{gg}^n)\,,
\end{align}
while $H_{\mathrm{GM}}$ transforms to zero. Notice that, due to the multi-frequency driving, it is not possible to find a reference frame where the Hamiltonian is time-independent. In spite of the time dependence, it is still possible to adiabatically eliminate the excited states. For this purposes, we define a projector operator for the atomic subspace, $\PP=\sum_n (\sigma_{gg}^n+\sigma_{ss}^n)$, that projects out the excited states, and its orthogonal counter part, $\QQ=\sum_n \sigma_{ee}^n$. Using these operators, one can formally project out slow and fast subspaces in the Schr\"odinger equation:
\begin{align}
 i \frac{d \PP \ket{\Psi}}{dt}&=\PP H\PP\ket{\Psi}+\PP H \QQ\ket{\Psi}\,,\nonumber \\
 i \frac{d \QQ \ket{\Psi}}{dt}&=\QQ H\PP\ket{\Psi}+\QQ H \QQ\ket{\Psi}\,.
\end{align}

By using the fact that $\QQ H \QQ$ is actually time-independent and assuming that initially there are no contributions from the excited states, i.e., $\QQ\ket{\Psi(0)}=0$, one can formally  integrate $\QQ \ket{\Psi}$ (by parts), input the result into the equation of $\PP\ket{\Psi}$, and obtain an effective Hamiltonian for the slow subspace:
\begin{align}
 i \frac{d \PP \ket{\Psi}}{dt}&\approx (\PP H\PP-\PP H \QQ\frac{1}{\QQ H \QQ}\QQ H\PP )\PP\ket{\Psi}\equiv H_{\mathrm{eff}}\PP\ket{\Psi}\,.
\end{align}
The resulting effective Hamiltonian then reads as the following:
\begin{align}
H_{\mathrm{eff}}&=\tilde{H}_{\mathrm{eff},\aa}+ \tilde{H}_{\mathrm{eff,lm}}+\tilde{H} \nonumber \\
&=\sum_{n}\big(\omega_{g,n}\sigma_{gg,n}-\frac{\Omega(t)\Omega^*(t)}{4\Delta}\sigma_{ss,n}\big) \nonumber \\
&-\sum_{\kk,n}\frac{g_\kk (\rr_n)\Omega(t)}{2 \Delta} a_\kk \sigma_{sg}^n e^{i (\omega_\mathrm{L}-\omega_\kk) t}+\mathrm{h.c.}\,,\nonumber \\
 &-\sum_{\kk,n}\frac{|g_\kk (\rr_n)|^2}{ \Delta} a^\dagger_\kk a_\kk \sigma_{gg}^n \,,
\end{align}
where we have absorbed some irrelevant phases. The contribution of $\tilde{H}=-\sum_{\kk,n}\frac{|g_\kk (\rr_n)|^2}{ \Delta} a^\dagger_\kk a_\kk \sigma_{gg}^n$ will be negligible since it is proportional to the number of GM photons, which is close to $0$ in our situation \cite{douglas15a, gonzaleztudela15c}. Rewriting the effective Hamiltonian in the interaction picture with respect to $\tilde{H}_{\mathrm{eff},\aa}$, we arrive at the following light-matter Hamiltonian
\begin{align}
H_{\mathrm{eff, lm}}(t)=-\sum_{\kk,n}\frac{g_\kk (\rr_n)\Omega(t)}{2 \Delta} a_\kk \sigma_{sg}^n e^{i (\omega_\mathrm{L}-\omega_\kk-\omega_{g,n}) t}+\mathrm{h.c.}\,. \label{eqSM:lmeff}
\end{align}
Note that, for simplicity in the derivation, we have neglected the contribution of the time-dependent Stark-Shift in the $\ket{s}_n$ states, which is given by
\begin{align}
 \delta\omega_{s}(t)&=-\frac{\Omega(t)\Omega^*(t)}{4\Delta} \nonumber \\
 &= -\sum_{\alpha=0}^{m_P-1}\frac{|\Omega_\alpha|^2}{4\Delta}-\sum_{\alpha>\beta}^{m_P-1} \Re \Big[\frac{\Omega_\alpha \Omega_{\beta}^*}{2\Delta} e^{i (\tilde{\omega}_\alpha-\tilde{\omega}_\beta)t} \Big]\,.
\end{align}
Here, $\Re[.]$ indicates real part. The time-independent contribution can be absorbed into the energy of $\omega_s$ without significant contribution to the dynamics, whereas the time-dependent terms will be averaged out in the atomic timescales that we are interested in. We consider its possible detrimental effects in the main text and in Appendix~\ref{secSM:error}, where we analyze the limitations and other error sources.

The relaxation timescales of the GMs in the PCWs are typicality much faster than the atomic ones, such that we can trace out the photonic information to obtain an effective master equation that describes the dynamics of the atomic system through its density matrix evolution \cite{gardiner_book00a}
\begin{align}\label{eqSM:masterE}
 \frac{d\rho}{dt}=\sum_{m,n}\Big[ &\Gamma_{m,n}(t)\big(\sigma_{sg}^n\rho\sigma_{gs}^m-\sigma_{sg}^n\sigma_{gs}^m\rho\big) \nonumber \\ 
 &+\Gamma_{m,n}^*(t)\big(\sigma_{sg}^m\rho\sigma_{gs}^n-\rho\sigma_{sg}^m\sigma_{gs}^n\big) \Big]\,,
\end{align}
where the time-dependent coefficients are given by
\begin{align}
 \Gamma_{m,n}(t)&=\nonumber \\
 &\int_0^\infty ds f_{\kk,m,n} e^{i(\omega_{g,m}-\omega_{g,n})t} e^{-i(\omega_{k}+\omega_{g,m}-\omega_{\mathrm{L}})s}\frac{\Omega(t)\Omega^*(t-s)}{4\Delta^2}
\end{align}
with $f_{\kk,m,n}=\sum_{\kk}|g_\kk|^2 e^{i\kk\cdot(\rr_n-\rr_m)}$. Expanding $\Omega(t)\Omega^*(t-s)=\sum_{\alpha,\beta}\Omega_{\alpha}\Omega_{\beta}^* e^{i(\tilde{\omega}_\alpha-\tilde{\omega}_\beta)t} e^{i \tilde{\omega}_\beta s}$, we find that:
\begin{align}
 \Gamma_{m,n}(t)&=\sum_{\alpha,\beta}\frac{\Omega_{\alpha}\Omega_{\beta}^*}{4\Delta^2}\Gamma_{m,n,\beta,\infty} e^{i(\omega_{g,m}-\omega_{g,n}+\tilde{\omega}_{\alpha}-\tilde{\omega}_{\beta})t}\,,
\end{align}
where $\Gamma_{m,n,\beta,\infty}$ is the time-independent contribution that can be written:
\begin{align}
 \Gamma_{m,n,\beta,\infty}&=\int_0^\infty ds f_{\kk,m,n} e^{-i(\omega_{k}+\omega_{g,m}-\omega_{\mathrm{L}}-\tilde{\omega}_\beta)s}\,.
\end{align}

Using that $\int_0^\infty e^{i x \tau} d\tau = \pi \delta(x)- i \mathrm{P}\frac{1}{x}$, and assuming that we are working in the regime within the bandgap [see Fig. \ref{fig1}(d)], i.e., $\Delta_{m,\beta}=\omega_c-\omega_{L}+\omega_{g,m}-\tilde{\omega}_\beta<0$ such that the dissipative terms [$\delta$-contribution] vanish, $\Gamma_{m,n}(t)$ contains only dispersive contributions,
\begin{align}\label{eqSM:timecoupling}
 \Gamma_{m,n}(t)=i \sum_{\alpha,\beta}\frac{\Omega_{\alpha}\Omega_{\beta}^*}{4\Delta^2} \tilde{J}_{m,n} e^{i(\omega_{g,m}-\omega_{g,n}+\tilde{\omega}_{\alpha}-\tilde{\omega}_{\beta})t}\,,
\end{align}
and $\tilde{J}_{m,n}$ is defined as
\begin{align}
\tilde{J}_{m,n} &= \sum_{\kk\in \mathrm{BZ}}  \frac{|g_k|^2}{ (\omega_{\mathrm{L}}-\omega_{k}-\omega_{g,m}+\tilde{\omega}_\beta)} e^{ i \kk \cdot (\rr_n - \rr_m)}\,,
\end{align} 
coinciding with the expressions obtained in Refs. \cite{douglas15a,gonzaleztudela15c} when $|\tilde{\omega}_\beta-\omega_{g,m}|\ll |\omega_L-\omega_c|$.  Then, depending on the dimensionality of the reservoir, we have \cite{douglas15a,gonzaleztudela15c}
\begin{align}\label{eqSM:1d}
\tilde{J}_{m,n,\mathrm{1d}} &\propto e^{-|\rr_{m,n}|/\xi}\,,\\
\tilde{J}_{m,n,\mathrm{2d}} &\propto K_0(|\rr_{m,n}|/\xi)\,,
\end{align} 
where $\rr_{m,n}=\rr_m-\rr_n$ and and $K_0(x)$ is the modified Bessel function of the second kind; $\xi=\sqrt{A/|\Delta_{\mathrm{xy}}|}$ controls the effective range of the interaction; $\Delta_{\mathrm{xy}}=|\omega_c-\omega_L|$ is the effective detuning with respect to the band edge, and $A$ the curvature of the band [see Fig. \ref{fig1}(c)]. Notice that there is another underlying assumption in the derivation, namely, that the coupling strength $|g_{\kk}|^2$ of the driven GMs must be approximately constant for all the sideband frequencies $\omega_{\mathrm{L}} + \tilde{\omega}_\alpha$. If not, the variation of $|g_\kk|^2$ can be compensated by adjusting sideband amplitudes $\Omega_\alpha$. Furthermore, since we focus on full control introduced by multi-frequency drivings, we will not specify the form of $\tilde{J}_{m,n}$ and simply assume a constant $\tilde{J}_{m,n}=\tilde{J}$ throughout the interaction range considered. However, one should be aware that the length scale $\xi$ will pose the ultimate 
limitation of the range of the interactions that we can simulate.

Since we have $\Gamma^*_{m,n}(t)=-\Gamma_{n,m}(t)$, the evolution of the density matrix in Eq.~(\ref{eqSM:masterE}) is governed by an effective XY Hamiltonian
\begin{align}
 \label{eqSM:XYham}
 H_{\mathrm{xy}}(t)=\sum_{m,n\neq m}^N\sum_{\alpha,\beta}\frac{\Omega_{\alpha}\Omega_{\beta}^*}{4\Delta^2} \tilde{J}_{m,n} e^{i(\omega_{g,m}-\omega_{g,n}+\tilde{\omega}_{\alpha}-\tilde{\omega}_{\beta})t}\sigma_{gs}^m\sigma_{sg}^n\,.
\end{align}
Interestingly, if we choose $\omega_{q,n}-\omega_{q,m}\neq 0$, we can control the resonant processes by adjusting the laser frequencies $\tilde{\omega}_{\alpha}$. In particular, two atoms $n$ and $m$, will be interacting through a resonant process with rate $\Omega_{\alpha}\Omega_{\beta}^* \tilde{J}_{m,n}/(4\Delta^2)$ when the resonant condition,
\begin{equation}
 \omega_{g,m}-\omega_{g,n}=\tilde{\omega}_{\beta}-\tilde{\omega}_{\alpha}\,,\label{eqSM:resonant}
\end{equation}
is satisfied. The intuitive picture is depicted in Fig. \ref{fig2} (a) in the main text: the atom $n$ scatters from sideband $\alpha$ a photon with energy $\omega_\LL+\tilde{\omega}_{\alpha}-\omega_{g,n}$ into GMs. When this photon propagates to atom $m$, it will only be absorbed via a sideband $\beta$ that satisfies $\omega_\LL+\tilde{\omega}_{\alpha}-\omega_{g,n}=\omega_\LL+\tilde{\omega}_{\beta}-\omega_{g,m}$, with the rest of the sidebands being off-resonant; Fig.~\ref{fig2} (b) depicts the reversed process. Therefore, the Hamiltonian $H_{\mathrm{xy}}(t)$ can be separated into time-independent (on-resonant) and time-independent (off-resonant) contributions: $H_{\mathrm{xy}}(t)=H_{\mathrm{xy},0}+\tilde{H}_{\mathrm{xy}}(t)$, where $H_{\mathrm{xy},0}$ is an XY spin Hamiltonian,
\begin{align}
 \label{eqSM:XYham2}
 H_{\mathrm{xy},0}=\sum_{m,n\neq m}^N J_{m,n}\sigma_{gs}^m\sigma_{sg}^n\,,
\end{align}
where the spin exchange coefficient $J_{m,n}$ can be fully controlled by adjusting $\Omega_\alpha$ and $\tilde{\omega}_\alpha$. We have
\begin{align}
 \label{eqSM:coupling222}
 J_{m,n}=\sum_{\alpha,\beta}\frac{\Omega_{\alpha}\Omega_{\beta}^*}{4\Delta^2} \tilde{J}_{m,n}\delta( \omega_{g,m}-\omega_{g,n}+\tilde{\omega}_{\alpha}-\tilde{\omega}_{\beta}) \,,
\end{align}
where $\delta(0)=1$ and $\delta(x\neq 0)=0$. Note that to fully control $J_{m,n}$ at each distance $\rr_{m,n}$, we need to introduce enough sidebands to cover all the energy differences $\omega_{g,m}-\omega_{g,n}$. 

If the characteristic energy scale of the spin Hamiltonian $H_{\mathrm{xy},0}$ is much smaller than the minimum energy detuning $\delta_\omega \equiv \min\{ |\omega_{g,n}-\omega_{g,m}|\}$ between different sites, that is $\delta_\omega \gg \Omega_{\alpha}\Omega_{\beta}/(4\Delta^2)$, the time-dependent processes will be highly off-resonant, yielding the ideal Hamiltonian $H_{\mathrm{xy}}(t)\approx H_{\mathrm{xy},0}$. In practical situations, $\delta_\omega$ will be a limited resource because of the requirement for large field gradient over the distance of a PCW unit cell. Thus, in the main text and in Appendix \ref{secSM:error}, we discuss errors created by the time-dependent processes, and strategies to minimize the errors.

As a last remark, if we explicitly write down the phase dependence of the Raman pump in Eq.~(\ref{eqSM:XYham2}), it follows that $J_{m,n}\equiv |J_{m,n}| e^{i \kk_\LL\cdot \rr_{m,n}}$ (and now $J_{m,n}=J_{n,m}$ for $\kk_\LL\cdot \rr_{m,n} = 0$). If the illumination is not perfectly transverse, that is, $\kk_L\cdot \rr_{m,n}\neq 0$, then $H_{\mathrm{xy},0}$ acquires spatial-dependent, complex coupling coefficients:
\begin{align}
 \label{eqSM:XYham3}
 H_{\mathrm{xy},0}=\sum_{m,n> m}^N J_{m,n}\big(e^{i \kk_\LL\cdot \rr_{m,n}}\sigma_{gs}^m\sigma_{sg}^n+e^{-i \kk_\LL\cdot \rr_{m,n}}\sigma_{gs}^n\sigma_{sg}^m\big)\,,
\end{align}
which gives us the possibility to engineer geometrical phases and non-trivial topological spin models. We also note that, in the above simple expression, we are assuming $\kk_\alpha = \kk_\LL$ being a constant for all different sidebands $\Omega_\alpha$.

\subsection*{Independent control of XX and YY interactions}

So far, we have been able to engineer full control of spin-exchange or $XY$ Hamiltonians. In this Section, we will show how by slight modification of the atomic level structure, we can engineer the $XX$ and $YY$ terms independently. In particular, we use a \emph{butterfly}-like structure as depicted in Fig.~\ref{fig2b}~(b), where there are two transitions coupled to the GMs, i.e., $\ket{g}\leftrightarrow \ket{e}$ and $\ket{s}\leftrightarrow \ket{\tilde{e}}$ and two different multi-frequency Raman fields, $\Omega_g(t)$ and $\Omega_s(t)$. Assuming that we have co-propagating beams or perfectly transverse illumination, i.e., $\kk_\LL \cdot \rr_{m,n} = 0$, we can adiabatically eliminate the excited states $\ket{e}$ and $\ket{\tilde{e}}$ following a similar procedure as in the previous Section [Eq~.(\ref{eqSM:lmeff})] to obtain an effective light-matter Hamiltonian
\begin{align}
H_{\mathrm{eff,lm}}(t) &= \nonumber \\
&- \sum_{\kk,n}\frac{g_\kk (\rr_n)\Omega(t)}{2 \Delta} a_\kk( \sigma_{sg}^n+e^{-i\phi_{gs}}\sigma_{gs}^n)e^{i (\omega_\mathrm{L}-\omega_\kk-\omega_{g,n}) t} \nonumber \\
&+\mathrm{h.c.}\,, \label{eqSM:lmme}
\end{align}
where we assumed that $\Omega_{s}(t)/\Delta_{s}=\Omega_{g}(t)e^{i\phi_{gs}}/\Delta_{g} \equiv\Omega(t)/\Delta$, $\Delta_{s,g}=\omega_e-\omega_{\LL,s,g}$, and $\phi_{gs}$ is the relative phase between the two multi-frequency Raman fields that can be adjusted at will. 

Adiabatically eliminating the photonic modes under all the approximation that we used in the previous Section, we arrive at an effective Hamiltonian
\begin{align}
 \label{eqSM:XXham3}
 H_{\mathrm{xx,yy,0}}=\sum_{m,n> m}^N \Big[J_{m,n}(\sigma_{gs}^m+e^{i\phi_{gs}}\sigma_{sg}^m)(\sigma_{sg}^n+e^{-i\phi_{gs}}\sigma_{gs}^n)+\mathrm{h.c.}\Big]\,,
\end{align}
which, depending on the phase $\phi_{gs}$, can drive either $X$ or $Y$ component, i.e., $(\sigma_{sg}^m \pm \sigma_{gs}^m)$, or more exotic combinations for general $\phi_{gs}$. Moreover, if the two pump beams are not co-propagating, they create spatial dependent phases $\phi_{gs}$. This can create site dependent $XX$, $YY$ or $XY$ terms.

\section{Proper choice of ground state energy shifts in 2D models}\label{secSM:Zeeman}
For generic 2D lattice models, we need to introduce ground state energy shifts between sites $(m, n)$ separated by $\rr_{m,n} = \rr_m - \rr_n$ to engineer the interaction between them. The energy shifts need to be unique for specific site separation vector $\rr_{m,n}$, but should be independent of $\rr_m$ to preserve translational invariance, which is generally required for spin lattice models. To do this, we can introduce a \emph{linear} magnetic field gradient $\mathbf{\nabla}B \cdot \mathbf{a}_l \equiv B_l$, where $\mathbf{a}_l$ ($l=1, 2$) are the Bravais vectors of a unit cell. We require that the ratio $q=B_{1}/B_{2}$ of B-field gradients along two Bravais vectors be an irrational number such that, for any $\rr_n=(n_x,n_y)$ with $n_x, n_y \in \mathbb{Z}$, $\omega_{g,n} =\mu_B (n_x\times q +n_y) B_2$ is a unique number. As a result, each sideband can only induce a resonant interaction at a specific separation $\rr_{m,n}$. Moreover,  we also need to ensure that there exists no $\rr_m$ such that $|\omega_{
g,m}- \omega_{g,n}|  \lesssim J_{m,n}$ that can lead to significant time-dependent terms in Eq.~(\ref{eq:Hamtk}) of the main text. In general, for a finite size system, this situation can be avoided.

\section{Pump field configurations for engineering a topological spin model in a brick-wall lattice}\label{secSM:brickwall}
In this Section, we describe in detail how to engineer Haldane's topological spin model Eq.~(\ref{eq:Haldane}) in the main text,
\begin{eqnarray}
H = t_1\sum_{\langle m,n\rangle} (\sigma_m^\dagger \sigma_n +\mathrm{h.c.} )+ t_2 \sum_{\{m,n\}} ( e^{i\phi_{mn}}\sigma_m^\dagger \sigma_n + \mathrm{h.c.}), \label{eqSM:Haldane}
\end{eqnarray}
in a brick-wall configuration. Introducing a strong pump field of amplitude $\Omega_0$, propagating along $\hat{y}$, we can engineer the interaction terms one-by-one as the following:

\begin{itemize}
 \item \emph{Uniform NN coupling along $\hat{y}$:} This term can be realized with a sideband of detuning $\delta_y$ and $X_y(\rr_n + \Delta\rr_y) = -\frac{|\Omega_1|}{2\Delta}e^{-i (n_y+1) \pi}$, pairing with the strong pump field $X_0(\rr_n) = \frac{|\Omega_0|}{2\Delta} e^{-i n_y\pi}$ which propagates along $\hat{y}$. The resulting coupling coefficient is $t_1 = \tilde{J} |X_0| |X_y|$. This term can also be extended to engineer non-uniform coupling coefficients; see the following discussion.
 
 \item \emph{Checkerboard-like NN coupling along $\hat{x}$:} We introduce a sideband of detuning $\delta_x$ and amplitude $X_x(\rr_n + \Delta \rr_x) = \frac{|\Omega|}{4\Delta} [  e^{-i  n_y \pi} + \zeta e^{-i  (n_x+1) \pi} ]$, formed by two fields propagating along $\hat{y}$ and $\hat{x}$, respectively. If both fields have the same amplitude ($\zeta =1$), they either add up or cancel completely depending on whether $n_x+n_y$ is odd or even. The resulting coupling rate is real with amplitude 
 \begin{equation}
  \label{eqSM:mmxx}
  J_{m,n} = \tilde{J} X_0 X_x^* = \frac{t_1}{2} \left[1- (-1)^{n_x - n_y} \right] \,,
 \end{equation}
and vanishes exactly in a checkerboard pattern. If one applies the same trick toward NN coupling along $\hat{y}$, but with $\zeta \neq 1$, the coupling amplitude also modulates in a checkerboard pattern. Essentially all three NN terms that from a brick-wall vertex can be independently controlled, opening up further possibility to engineer, for example, Kitaev's honeycomb lattice model \cite{kitaev2006,feng07}. 

\item \emph{Complex NNN and NNNN couplings:} The NNNN terms can be similarly generated by using a sideband of detuning $2\delta_{y}$, and $X_{2y}(\rr_n+\Delta \rr_{2y}) = \frac{|\Omega_2|}{2\Delta}[  e^{-i  (n_y+2) \pi} + i \zeta e^{-i  n_x \pi}]$, formed by two initially $\pi/2$ out-of-phase fields propagating respectively along $\hat{y}$ and $\hat{x}$. The same trick can be used for NNN couplings using sidebands of detunings $\delta_{xy},\delta_{xy^*}$ respectively. The coupling phase $\phi$, can be arbitrarily controlled by the amplitude ratio $\zeta$, as we showed in Eq.~(\ref{eq:mm}) in the main text.
\end{itemize}

To engineer this model, again only two pump beams can introduce all components required in the Raman field. We can write down the $\hat{x}$-, $\hat{y}$-propagating fields with amplitude modulations (that is, with equal blue- and red-sideband contributions), for the field propagating along $\hat{y}$, we have
\begin{align}
\epsilon_y(t) =  \epsilon_0 & + \epsilon_1  \left( \frac{1}{2} \cos \delta_x t - \cos \delta_y t\right)   \nonumber\\
&+  \epsilon_2 \left(\cos2 \delta_{y} t - \cos \delta_{xy}t - \cos \delta_{xy^*} t \right), \label{eqSM:eybrick}
\end{align}
and, for the field propagating along $\hat{x}$, we require
\begin{align}
\epsilon_x(t) =  \frac{\epsilon_1}{2} \cos \delta_x t  -i \zeta \epsilon_2 \left( \cos 2 \delta_{y} t + \cos \delta_{xy}t - \cos \delta_{xy*} t \right). \label{eqSM:exbrick}
\end{align}
One may similarly replace amplitude modulations $\cos\delta_\alpha t$ by frequency modulation $e^{- i \delta_\alpha t}$ to engineer the spin model, as in Eqs.~(\ref{eq:eychiral}-\ref{eq:exchiral}) in the main text.

\section{PCW and pump field configurations for engineering a XXZ spin Hamiltonian with $1/r^\eta$ interaction} \label{secSM:Hxxz}
We describe how to engineer a XXZ Hamiltonian, which is typically written as
\begin{align}
 \label{eqSM:spin}
H_{XXZ} & = -B \sum_{n} \sigma_n^z \nonumber \\
&+\sum_{n<m} \frac{J}{|\rr_{n}-\rr_{m}|^\eta}
\Big[\cos(\theta)\sigma_n^z\sigma_m^z+\sin(\theta)(\sigma_n^x\sigma_m^x+\sigma_n^y\sigma_m^y)\Big]\,,
\end{align}
\noindent where an effective magnetic field $B$ that controls the number of excitations can be introduced by the detuning of addressing beams; The parameter $\theta$ determines the relative strength between the $ZZ$ and $XY$ interactions. 

We can simulate any $\eta$ and $\theta$ in a 2D PCW as follows:

\begin{itemize}
\item First, we require the decay length scale $\xi$ to be sufficiently large such that GM photon coupling strength $\tilde{J}(\rr_{m,n})$ is only limited by the energy spread in the low dimensional reservoir. For example, in 2D with quadratic dispersion as depicted in Fig. \ref{fig1}(d) of the main text, one can simulate any interaction that decays faster than $ K_0(|\rr_{m,n}|/\xi)\approx \log(\xi/|\rr_{m,n}|)$, where $K_0(x)$ is the modified Bessel function of the second kind \cite{douglas15a,gonzaleztudela15c} (Appendix~\ref{secSM:derivation}).

\item Then, in order to simulate the $|\rr_{n}-\rr_{m}|^{-\eta}$ dependence, we introduce a linear magnetic field gradient $\mathbf{\nabla}B$ or linear ground state energy shifts as described in Appendix~\ref{secSM:Zeeman}.

\item Thus, as we did in the previous discussions, we introduce a strong pump field of amplitude $\Omega_{0}$ together with $N_d$ auxiliary fields $\Omega_\alpha$ of detuning $\delta_\alpha$ to cover all the different separations $\rr_{m,n}$. For example, to simulate a square lattice of $n_s\times n_s$ ($=N$) atomic spins, we can see that the number of different distances grows as $N_d=\frac{n_s(n_s+1)-2}{2}$, linearly proportional to the number of atoms $N_d\propto N$. 

\item Finally, the parameter $\theta$ that determines the ratio between $ZZ$ and $XY$ interaction can be controlled by using different pump intensities when doing the stroboscopic evolution (see Section \ref{sec:ZZ}).

\end{itemize}

\section{Engineering a two-step Hamiltonian for error reduction}\label{secSM:two-step}
As derived in Section \ref{secSM:derivation}, the resulting time-dependent couplings introduced by a multi-frequency pump is $H(t)=\sum_p H_p e^{i p \delta t}$, where $H_p$ represents the part in the Hamiltonian that oscillates with frequency $p\delta$. The time-dependent Hamiltonian has a period $T=2\pi/\delta$, and it can be shown that the effective Hamiltonian that repeats every time-period $T$, i.e., $e^{-i H_{\mathrm{eff}}T}$, where the measurements would take place in an experiment, is given by \cite{goldman14a}
\begin{align}
\label{eqSM:effhamflo}
H_{\mathrm{eff,1}} &\approx  H_0+ \frac{1}{\delta}\sum_p \frac{[H_p,H_{-p}]}{p} \nonumber \\
 &+ \frac{1}{2\delta^2}\sum_p\frac{[[H_p,H_0],H_{-p}]+[[H_{-p},H_0],H_{p}]}{p^2}\,.
\end{align}
This means that if we apply the multi-frequency pumps just as we explained in the previous sections, the leading error would be on the order of $J^2/\delta$, where $J$ is the interaction strength that we want to simulate. However, we also observe that the leading error term $\sum_p [H_p,H_{-p}]/(p\delta)$ vanishes if $H_p = \pm H_{-p}$. In other words, first order error vanishes if $H_p$ is either symmetric or anti-symmetric under time-reversal operation $\mathcal{T}$. While the original Hamiltonian $H(t)$ doesn't necessarily posses such symmetry, it is possible to introduce a two-step periodic operation $H_\mathrm{2step} = \{ H, \mathcal{T} H, H, \mathcal{T} H , ...\}$ to cancel the first order error while keeping the time-independent part $H_\mathrm{2step,0}=H_0$ identical. This reduces the leading error to the order of $J^3/\delta^2$. 

To achieve the time-reversal operation, we must reverse the phase of the driving lasers, as well as the sign of the energy offsets between the atoms. Specifically, we can engineer the two-step Hamiltonian as follows:
\begin{itemize}
\item After applying a proper magnetic (or Stark-shift) gradient to ensure a position dependent energy shift, we apply sidebands using either frequency or amplitude modulations to engineer the Hamiltonian $H_0$ following Eq.~(\ref{eqSM:XYham2}).

\item After a time $T=2\pi/\delta$, we flip the sign of the gradient. In 1D, if $\omega_{g,n}-\omega_{g,m}=|n-m|\delta$, then we switch to $\omega_{g,n}-\omega_{g,m}=-|n-m|\delta$. Meanwhile, we also reverse the propagation direction of the Raman fields such that $X_\alpha\rightarrow X_{\alpha}^*$. 
As a result, all the time-dependent Hamiltonians $H_{p}$, $\forall p \neq 0$, become $H_{-p}$ in the second step. Whereas, the time-independent Hamiltonian $H_0$ remains identical. After holding for an evolution time $T=2\pi/\delta$, we repeat the first step.

\end{itemize} 

To formally prove that the above idea leads to smaller error, we can write our two step periodic Hamiltonian as follows:
\begin{equation}
H_\mathrm{2step}(t) = H (t) T(t) + H (-t) T(t-T), \label{eqSM:twostep}
\end{equation}
where $T(t)$ is a periodic square-wave envelope, controlling the on and off of $H(t)$ at time interval $[0,T]$ within a period $2T$. $T(t)$ can be expanded as follows
 \begin{eqnarray}
T(t) &=&  \frac{1}{2} + \frac{1}{\pi i} \sum_{m~\mathrm{odd}} \frac{(e^{i m \delta t/2}-e^{-i m \delta t /2})}{m}\,,
\end{eqnarray}
where $m=1,3,\dots$. Plugging the expansion $H(t) = \sum_p H_{p} e^{ip\delta t}$ into Eq.~(\ref{eqSM:twostep}), we now have
\begin{align}
  H_\mathrm{2step}(t) &= \frac{1}{2} \sum_p \Big[ H_{p} e^{i p \delta t} \big(1 + \frac{2}{ \pi i} \sum_{m~\mathrm{odd}} \frac{(e^{i m \delta t/2}-e^{-i m \delta t/2})}{m} \big)   \nonumber\\  &+ H_{p  } e^{-i p \delta t} \big(1 - \frac{2}{\pi i} \sum_{m~\mathrm{odd}} \frac{(e^{i m \delta t/2}-e^{-i m \delta t/2})}{m} \big)\Big]\,. 
\end{align}

Writing $H_\mathrm{2step}(t) = \sum_p \tilde{H}_p e^{ i p (\delta/2) t}$, the Fourier components $\tilde{H}_p$ of the two-step periodic Hamiltonian are
\begin{eqnarray}
\tilde{H}_{p(\mathrm{even})} &=& \frac{1}{2} (H_{p/2}+H_{-p/2})\,, \label{eqSM:even} \\
\tilde{H}_{p(\mathrm{odd})} &=&  \frac{1}{\pi i} \sum_{m~\mathrm{odd}} \frac{1}{m} (H_{(p-m)/2} - H_{-(p-m)/2} \nonumber \\
&+& H_{(p+m)/2} - H_{-(p+m)/2}\big)\,.\label{eqSM:odd}
\end{eqnarray}

So we have $\tilde{H}_p = (-1)^p\tilde{H}_{-p}$  and the leading order error in Eq~(\ref{eqSM:effhamflo}) vanishes. According to the Floquet theory, we arrive at an effective time-independent Hamiltonian $H_\mathrm{eff,2}$ at every time interval $T_2=4\pi/\delta$,
\begin{align}
\label{eqSM:error}
H_{\mathrm{eff,2}} = H_0+H_{\mathrm{err,2}} \approx H_0+ \frac{4}{\delta^2}\sum_p(-1)^p\frac{[[\tilde{H}_p,H_0],\tilde{H}_{p}]}{p^2}\,.
\end{align}

\section{The role of time-dependent Stark shifts in the \emph{butterfly} scheme}\label{secSM:error}

We discuss the error contribution due to time-dependent Stark shifts in the \emph{butterfly} scheme for independent control of XX or YY interactions (as well as ZZ interaction in stroboscopic evolution). The time-dependent Stark shift is
\begin{align}
\label{SMeq:acshift}
H_{\mathrm{ac}}(t)= &-\sum_n\sum_{\alpha>\beta}^{m_P-1} \Re[\frac{\Omega_{g,\alpha}\Omega^*_{g,\beta}}{2\Delta_g} e^{i\tilde{\omega}_{\alpha,\beta}t} ]\sigma_{gg}^n \nonumber \\
&- \Re[\frac{\Omega_{s,\alpha}\Omega^*_{s,\beta}}{2\Delta_s} e^{i\tilde{\omega}_{\alpha,\beta}t} ]\sigma_{ss}^n\,,
\end{align}
where $\tilde{\omega}_{\alpha,\beta}=\tilde{\omega}_\alpha-\tilde{\omega}_\beta$. The Fourier coefficients of the Stark shifts can be written as
\begin{align}
H_{\mathrm{ac},p} = \sum_n \frac{A_p^n}{\Delta} \big( \Delta_g \sigma_{gg}^n + \Delta_s \sigma_{ss}^n \big),
\end{align}
where $A_p^n = - \sum_{\alpha \neq \beta} \frac{\Omega_\alpha \Omega_\beta^*}{4 \Delta} \delta(\tilde{\omega}_{\alpha,\beta} - p \delta) $ and we have used that fact that $|\Omega_{g,\alpha}|/\Delta_g = |\Omega_{s,\alpha}|/\Delta_s = |\Omega_\alpha|/\Delta$. In two-step driving, we replace $A_p^n$ with $\tilde{A}_p^n$ according to Eqs.~(\ref{eqSM:even}-\ref{eqSM:odd}). We note that states $\ket{g}$ and $\ket{s}$ can be driven differently when $\Delta_g \neq \Delta_s$. This leads to different error comparing to simple Raman driving only on one of the state.

In two-step driving, the leading Stark-shift error contribution appears in the second order
\begin{align}
\label{SMeq:acerror}
H_{\mathrm{err,2}} \approx \frac{4}{\delta^2}\sum_p(-1)^p\frac{[[\tilde{H}_p + \tilde{H}_{\mathrm{ac},p},H_0],\tilde{H}_{p} +\tilde{H}_{\mathrm{ac},p}]}{p^2},
\end{align}
where $\tilde{H}_{\mathrm{ac},p}$ are the Fourier coefficients of the two-step Stark-shift Hamiltonian. To simplify the discussion, we discuss XX interaction with $\phi_{gs} = 0$ and  
\begin{align}
H_{0} = H_{XX,0} = \sum_{m,n> m}^N J_{m,n}(\sigma_{gs}^m+\sigma_{sg}^m)(\sigma_{sg}^n+\sigma_{gs}^n).
\end{align}
For illustration, we evaluate the following nested commutators, $[[\tilde{H}_{\mathrm{ac},p},H_0], \tilde{H}_{\mathrm{ac},p}]$, to access the error contribution due to time-dependent Stark shifts. We find 

\begin{widetext}
\begin{align}
[[\tilde{H}_{\mathrm{ac},p},H_0],\tilde{H}_{\mathrm{ac},p}] & =- \sum_{m,n> m}^N \frac{J_{m,n}}{\Delta^2}  ( \Delta_g -\Delta_s \big)^2 \big[ (\tilde{A}_p^m+\tilde{A}_p^n)^2 \big( \sigma_{gs}^m\sigma_{gs}^n +  \sigma_{sg}^m\sigma_{sg}^n \big) + (\tilde{A}_p^m-\tilde{A}_p^n)^2 \big( \sigma_{sg}^m\sigma_{gs}^n +  \sigma_{gs}^m\sigma_{sg}^n \big) \big] \\
&=- \sum_{m,n> m}^N \frac{J_{m,n}}{\Delta^2}  ( \Delta_g -\Delta_s \big)^2 \big\{ \big[ (\tilde{A}_p^m)^2+(\tilde{A}_p^n)^2 \big] \sigma_{x}^m\sigma_{x}^n  + 2\tilde{A}_p^m\tilde{A}_p^n \sigma_{y}^m\sigma_{y}^n  \big\}\,. \label{eqSM:acXXerror}
\end{align}
\end{widetext}
\noindent We therefore see that as long as $\Delta_g = \Delta_s$, there is no error due to time-dependent Stark shifts as a result that both $\ket{g}$ and $\ket{s}$ states are shifted exactly the same way. However, one may find that the atomic level structure for a butterfly scheme dictates that $\Delta_g \neq \Delta_s$. As a result, we find $[[\tilde{H}_{\mathrm{ac},p},H_0],\tilde{H}_{\mathrm{ac},p}] \neq 0$ even for translational-invariant models, where $\tilde{A}_p^n \equiv \tilde{A}_p$ are independent of sites. This is in contrast to the case of $XY$ Hamiltonians with the Raman field driving only one ground state. It is however still possible to minimize the error contribution by driving with $\tilde{J} > \Delta$ and $\delta \gg J$, as the criterion needed for sub-lattice models; see main text.

In the case of strong driving with $\Delta_g \neq \Delta_s$ and $\Delta > \tilde{J}$, the commutator $[[\tilde{H}_{\mathrm{ac},p},H_0],\tilde{H}_{\mathrm{ac},p}]$ can be made as the dominant `error' contribution. Interestingly, as seen from Eq.~(\ref{eqSM:acXXerror}), this `error' term is in fact also driving XX and YY interactions. It is therefore desirable to take the Stark shift contribution into account while finding the proper pump fields to create the target Hamiltonians.

\end{document}